\documentclass[12pt]{article}
\usepackage{graphics}
\usepackage[dvips]{graphicx}
\usepackage{mathrsfs}
\usepackage{amssymb}
\usepackage{verbatim}
\usepackage{float}
\newcommand{\be}{\begin{equation}}
\newcommand{\ee}{\end{equation}}
\newcommand{\bea}{\begin{eqnarray}}
\newcommand{\eea}{\end{eqnarray}}

\newcommand{\NPB}[3]{\emph{ Nucl.~Phys.} \textbf{B#1} (#2) #3}   
\newcommand{\PLB}[3]{\emph{ Phys.~Lett.} \textbf{B#1} (#2) #3}   
\newcommand{\PRD}[3]{\emph{ Phys.~Rev.} \textbf{D#1} (#2) #3}

\newcommand{\HPA}[3]{\emph{ Helv.~Phys.~Acta} \textbf{#1} (#2) #3}

\def\4vol{{\int d^4x \sqrt{-g}}}

\def\simlt{\stackrel{<}{{}_\sim}}

\newcommand{\nc}{\newcommand}

\nc{\nt}{\tilde{N}}
\nc{\ra}{\rightarrow}
\nc{\lsim}{\begin{array}{c}\,\sim\vspace{-21pt}\\< \end{array}}
\nc{\gsim}{\begin{array}{c}\sim\vspace{-21pt}\\> \end{array}}
\nc{\tnt}{\tilde{N}}
\nc{\tst}{\tilde{t}}

\nc{\LL}{L}
\nc{\vv}{\tilde{v}}

\title{
\vspace*{-1.3cm}
\begin{flushright}
\normalsize{
ANL-HEP-PR-04-137\\
EFI-04-40\\
FERMILAB-PUB-04-397-T
}
\end{flushright}
\vspace{1.5cm}
\Large
\textbf{The Supersymmetric Origin of Matter}
\vspace*{1.0cm}
\author{\large \textbf{C.~Bal\'azs$^{a}$}, \textbf{M.~Carena$^{b}$},
\textbf{A.~Menon$^{a,c}$}, \\
~\\
\textbf{D.E.~Morrissey$^{a,c}$}
and \textbf{C.E.M.~Wagner$^{a,c}$}\\ \\[0.5cm]
$^a$\normalsize\emph{HEP Division, Argonne National Laboratory,
9700 Cass Ave.,
Argonne, IL 60439, USA} \\
$^b$\normalsize\emph{Fermi National Accelerator Laboratory,
P.O.~Box 500, Batavia, IL 60510, USA}\\
$^c$\normalsize\emph{Enrico Fermi Institute, Univ. of Chicago,
5640 S. Ellis Ave., Chicago, IL 60637, USA}}} 

\begin{document}
\setcounter{page}{0}
\maketitle

\vspace*{1cm}
\begin{abstract}
The Minimal Supersymmetric extension of the Standard Model~(MSSM)
can provide the correct neutralino relic abundance and 
baryon number asymmetry of the universe.
Both may be efficiently generated in the presence of 
CP violating phases, light charginos and neutralinos, and a light 
top squark.  Due to the coannihilation of the neutralino with the light
stop, we find a large region of parameter space in which the
neutralino relic density is consistent with WMAP and SDSS data.
We perform a detailed study of the additional constraints induced 
when CP violating phases, consistent with the ones required for
baryogenesis, are included.
We explore the possible tests of this scenario from  present and
future electron Electric Dipole Moment~(EDM) measurements, direct
neutralino detection experiments, collider searches and the 
$b \to s \gamma$ decay rate.
We find that the EDM constraints are quite severe
and that electron EDM experiments, together with stop searches 
at the Tevatron and Higgs searches at the LHC,
will provide a definite test of our scenario of electroweak
baryogenesis in the next few years.
\end{abstract}

\thispagestyle{empty}
\newpage

\setcounter{page}{1}

\section{Introduction}

  The nature of the dark matter and the source of the baryon--anti-baryon 
asymmetry are two of the most important questions at the interface of 
particle physics and cosmology.  Recent improvements in the
astrophysical and cosmological data, most notably due to 
the Wilkinson Microwave Anisotropy Probe~(WMAP)~\cite{Spergel:2003cb}
and the Sloan Digital Sky Survey~(SDSS)~\cite{Tegmark:2003ud}, 
have determined the matter and baryon densities of the Universe 
to be $\Omega_Mh^2 = 0.135^{+0.008}_{-0.009}$ and 
\be
\Omega_Bh^2 = 0.0224\pm{0.0009}, 
\ee
respectively, 
with $h = 0.71^{+0.04}_{-0.03}$.  Together these imply
a (dominantly) cold dark matter density of 
\be
\Omega_{CDM}h^2 = 0.1126^{+0.0161}_{-0.0181},
\label{odm}
\ee
at $95\%$~CL.  Such precise determinations of $\Omega_Bh^2$
and $\Omega_{CDM}h^2$ impose severe constraints on any
particle physics model that tries to explain one or 
both of these values.

  The Standard Model of particle physics~(SM) has been tested
extensively by collider experiments, and has so far 
withstood all of them.  However, the SM performs considerably
worse when it comes to cosmology, and can account
for neither the baryon asymmetry, nor the dark matter.  
Furthermore, in the SM, the electroweak scale is unstable
under quantum corrections suggesting that an extension of
the SM description is required at energies near the~TeV~scale. 
A particularly attractive way to stabilize the weak scale is 
to introduce supersymmetry~\cite{Haber:1984rc}.  
Remarkably, the minimal supersymmetric 
extension of the standard model, the MSSM, can also explain
the baryon asymmetry, and contains an excellent dark matter
candidate in the lightest supersymmetric particle~(LSP).  

  The LSP of the MSSM is stable if R-parity is imposed.
If, in addition, the LSP is neutral under $SU(3)_C\times U(1)_{EM}$, 
it is a candidate for cold dark matter. 
One such particle is the lightest neutralino. 
This particle tends to have a mass of order of the weak scale and electroweak
strength couplings, and therefore naturally gives
rise to a dark matter relic density close to the measured value.
The fact that a stable particle with electroweak 
strength couplings and mass of order 1~TeV naturally generates 
a relic density near the required value can be taken as 
further motivation for new physics at the TeV scale. 

  In general, any mechanism for baryogenesis must fulfill the 
three Sakharov requirements~\cite{Sakharov:1967dj}; 
namely baryon number~(B) violation, CP violation, and a departure from equilibrium 
(unless CPT is violated, see for instance~\cite{Carmona:2004xc}).  
All three requirements are satisfied in both the SM and the 
MSSM during the electroweak phase transition, and this is the 
basis for electroweak baryogenesis~(EWBG)~\cite{EWBGreviews}.
However, as we will discuss below, while electroweak baryogenesis
may be realized in the MSSM, SM processes cannot generate
a large  enough baryon asymmetry during the electroweak phase transition.
  
  Baryon number violation occurs in the SM and the MSSM
due to anomalous sphaleron transitions that violate 
$(B\!+\!L)$~\cite{anomaly}.
These transitions are exponentially suppressed at low temperatures
in the electroweak broken phase~\cite{sphalerons}, 
but become active at high temperatures when the electroweak 
symmetry is restored~\cite{sphalT}.   
In the absence of other charge asymmetries, like $(B\!-\!L)$,
they produce baryons and anti-baryons such that the net
baryon number relaxes to zero, and so do not by themselves 
generate a baryon asymmetry~\cite{EWBGreviews,ckn}

  If the electroweak phase transition is first order, bubbles
of broken phase nucleate within the symmetric phase as the 
Universe cools below the critical temperature. These provide
the necessary departure from equilibrium.  
EWBG then proceeds as follows~\cite{Huet:1995sh}.
CP violating interactions in the bubble walls generate chiral 
charge asymmetries which diffuse into the symmetric phase in
front of the walls.  There, sphaleron transitions, 
which are active in the symmetric phase, convert these asymmetries 
into a net baryon number.  This baryon number then diffuses
into the bubbles where the electroweak symmetry is broken.
Sphaleron transitions within the broken phase 
tend to destroy the baryon number generated outside 
the bubble.  To avoid this, the sphaleron transitions within 
the broken phase must be strongly suppressed.  This is the case
provided the electroweak phase transition is \emph{strongly}
first order~\cite{Bochkarev:1987wf},
\be
v(T_c)/T_c \gtrsim 1\;,
\ee
where $v(T_c)$ denotes the Higgs vacuum expectation value
at the critical temperature $T_c$.  

  The strength of the electroweak phase transition may be 
determined by studying the finite temperature effective 
Higgs boson potential.  The Higgs vacuum expectation value at 
the critical temperature is inversely proportional to the
Higgs quartic coupling, related to the Higgs mass.  For
sufficiently light Higgs bosons, a first-order phase transition
can be induced by the loop effects of light bosonic particles,
with masses of order the weak scale and large couplings to the Higgs fields.  
The only such particles in the SM are the gauge bosons, 
and their couplings are not strong enough to induce
a first-order phase transition for a Higgs mass above the 
LEP~II bound~\cite{SMpt}.

  Within the MSSM, there are additional bosonic degrees
of freedom which can make the phase transition more strongly
first-order.  The most important contribution comes from
a light stop, which interacts with the Higgs field
with a coupling equal to the top-quark Yukawa.
In addition, a light stop has six degrees of freedom,
three of colour and two of charge, which further enhances the 
effect on the Higgs potential.  Detailed calculations show
that for the mechanism of electroweak baryogenesis to work,
the lightest stop mass must be less than the top mass
but greater than about 120~GeV to avoid colour-breaking minima.
Simultaneously, the Higgs boson involved in breaking the electroweak
symmetry must be lighter than 120~GeV~\cite{CQW}-\cite{LR},
only slightly above the present experimental bound~\cite{Barate:2003sz},
\be
m_h \gtrsim 114~\mbox{GeV}, 
\label{mhiggs}
\ee
which is valid for a Higgs boson with SM-like couplings 
to the gauge bosons.\footnote{The requirements of a light
stop and a light Higgs boson may be relaxed in non-minimal
supersymmetric extensions. See, for instance,
Refs.~\cite{Pietroni:in}--\cite{Carena:2004ha}.}

  The combined requirements of a first-order
electroweak phase transition, strong enough for EWBG,
and a Higgs boson mass above the experimental limit severely 
restrict the allowed values of the stop parameters.  
To avoid generating too large a contribution
to $\Delta\rho$, the light stop must be mostly right-handed.
Since the stops generate the most important radiative contribution
to the Higgs boson mass in the MSSM~\cite{Heinemeyer:1998jw}, 
the other stop must be considerably heavier in order to raise the Higgs mass
above the experimental bound, Eq.~(\ref{mhiggs}).
For the stop soft supersymmetry-breaking masses, this implies~\cite{Carena:1997ki}
\bea
m_{U_3}^2 &\lesssim& 0\;,\\ 
m_{Q_3}^2 &\gtrsim& (1~\mbox{TeV})^2\;.\nonumber
\eea
A similar tension exists for the combination of soft SUSY breaking
parameters defining the stop mixing,  
$|A_t-\mu^*/\tan{\beta}|/m_{Q_3}$, and $\tan\beta$. 
Large values of these quantities tend to increase the Higgs mass 
at the expense of weakening the phase transition or 
the amount of baryon number produced.  
The allowed ranges have been found to be~\cite{Carena:1997ki}
\bea
5&\lesssim &\tan{\beta}\:\: \lesssim \:\:10\;,\\ 
0.3&\lesssim &|A_t-\mu^*/\tan{\beta}|/m_{Q_3}\:\: 
\lesssim \:\:0.5\;.\nonumber
\eea

  A strong electroweak phase transition is only a necessary
condition for successful EWBG.  In addition, a CP violating
source is needed to generate a chiral charge asymmetry 
in the bubble walls.  Within the MSSM, the dominant source is
produced by the charginos, and is proportional to 
$Im(\mu\,M_2)$~\cite{improved,Carena:2002ss}.
For this source to be significant, the charginos must
be abundant in the plasma, which requires that they not be
too much heavier than the temperature of the plasma, $T\sim T_c$.
In the recent analysis of Ref.~\cite{Carena:2002ss}, the authors 
found the bounds
\bea
\label{Eq:phz}
|Arg(\mu\,M_2)| &\gtrsim& 0.1\;,\\
\mu,\:M_2 &\lesssim & 500~\mbox{GeV}\;.\nonumber
\eea
These conditions are very relevant to the issue of neutralino 
dark matter.

  The need for a large CP violating phase, Eq.~(\ref{Eq:phz}),
implies that there is a danger of violating the experimental 
bounds on the electric dipole moments~(EDM) of the electron,
neutron, and $^{199}$Hg atom since phases generate new
contributions to these EDM's.  The leading contributions arise
at one loop order, and they all contain an intermediate first or second
generation sfermion.  They become negligible if these sfermions are
very heavy, $m_{\tilde{f}}\gtrsim 10$~TeV.
Such large masses have only a very small effect on EWBG.
At two-loop order, if $Arg(\mu\,M_2)\neq 0$, there is a contribution 
involving an intermediate chargino and Higgs boson~\cite{Chang:2002ex,
Pilaftsis:2002fe}.  Since EWBG requires that this phase be non-zero and
that the charginos be fairly light, the two-loop contribution
is unavoidable if EWBG is to be successful.  
Thus, EDM limits strongly constrain the EWBG mechanism in the MSSM.  
Similarly, the branching ratio for $b\to s\gamma$ decays is also sensitive to
this phase, and therefore imposes a further constraint on the EWBG mechanism.


  In a previous work~\cite{Balazs:2004bu}, some of the present authors
investigated the neutralino relic density in the presence of 
a light squark, as required for EWBG, but without including the
effects of CP violating phases in the calculations.
Here, we extend the analysis to study in detail the effect of 
phases in order to better understand the relationship 
between EWBG and dark matter within the MSSM.
The outline of the paper is as follows.  In Section~\ref{ndm}
we investigate the relic density of a neutralino LSP
in the presence of both a light stop and CP violating phases.  
Section~\ref{direct} examines the prospects for direct detection of 
the neutralino dark matter in laboratory experiments, again
including CP violating phases.  In Section~\ref{edm}, we will look at 
the constraints on the phases needed for EWBG due to 
the electron EDM and flavour-violating $b\to\,s\gamma$ transitions. 
Finally, Section~\ref{concl} is reserved for our conclusions.

\section{Neutralino Dark Matter\label{ndm}}

  As discussed in the introduction, the dual requirements of 
successful EWBG and a lightest Higgs boson with mass greater 
than the LEP~II bounds strongly constrain the parameter space
of the MSSM \cite{Davidson:1999ii}.  
One of the stops must be light, with mass less
than that of the top, and mostly right-handed.  Furthermore,
the charginos must not be too heavy, and the combination
$\mu\,M_2$ must have a non-negligible phase.  These conditions 
have important implications for neutralino dark matter.  

  First of all, if the lightest neutralino is to be the 
source of the observed dark matter, it must be lighter
than the light stop so that it be stable.  Secondly,
in much of the parameter space of interest the light stop
is only slightly heavier than the neutralino LSP
implying that stop-neutralino coannihilation is significant.
Finally, a phase for $\mu\,M_2$ modifies the masses of the 
neutralinos and their couplings to other particles, and can
also affect the relative phase between the various contributions
to the annihilation cross-section.  
The effect of CP violating phases on neutralino dark matter has been
considered previously by several groups~\cite{Falk:1995fk}-
\cite{Argyrou:2004cs}.  However, in all of these analyses 
the regions of MSSM parameter space considered were much 
different from the restricted subset required for EWBG, 
and in particular, none of them included a light stop.  
 
  To simplify the analysis, we shall assume throughout this work
that the gaugino mass parameters $M_1$ and $M_2$ are 
related by the standard unification relation, 
$M_2 = (g_2^2/g_1^2)\,M_1\simeq 2\,M_1$.
The stop soft parameters are largely 
fixed by the EWBG and Higgs mass conditions.
We take them to be
\bea
m_{U_3}^2 &\approx& 0 \nonumber \\
m_{Q_3} &=& 1.5~\mbox{TeV} \\
|X_t| &=& |A_t - \mu^*/\tan\beta| \:\:=\:\: 0.7~\mbox{TeV}.\nonumber
\label{stopsoft}
\eea
We also set $m_{D_3} = m_{L_3} = m_{E_3} = 1$~TeV.
EWBG and the Higgs mass constraint also require $5\lesssim 
\tan\beta\lesssim 10$ and $M_A \gtrsim 200$~GeV.
For concreteness, we shall consider the values
\bea
\tan\beta &=& 7\\
M_A &=& 200,~1000~\mbox{GeV}.\nonumber
\label{higgspara}
\eea
The first and second generation sfermion soft masses are 
taken to be very large, $m_{\tilde{f}} \gtrsim 10$~TeV.  
As we will discuss in section~\ref{edm}, this is necessary 
to avoid the electron, neutron, and $^{199}$Hg EDM constraints in the 
presence of large phases.  

  The only phase that we consider in this work is the one
directly related to EWBG, namely $Arg(\mu\,M_2)$.  
We will assume further that this 
phase is the result of a common phase for the gaugino mass parameters.
With this assumption, all CP violating effects are confined to the 
chargino and neutralino sectors, or the loop corrections induced by 
them.\footnote{We do not consider the effects of a gluino phase.  For the 
parameters considered in the present work, we expect that such a phase would 
only have a very small effect.}
By means of a U(1)$_R$ transformation, we may transfer the gaugino phase 
into the $\mu$ parameter and the trilinear $A_f$ terms.  
Under this transformation, the effective values of these parameters 
are shifted according to
\bea
M_{\lambda} &\to& M_{\lambda}\;e^{-i\,\varphi},\\
\mu &\to& \mu\;e^{i\,\varphi},\nonumber\\
A_f &\to& A_f\;e^{-i\,\varphi},\nonumber
\label{Eq:u1r}
\eea
with the remaining MSSM parameters left unchanged.
For consistency of notation with~\cite{Carena:2002ss}, 
we will implicitly make a U(1)$_R$ rotation such that the gaugino masses are
all real and positive, and the $\mu$ parameter and the $A_f$ terms
have equal and opposite phases (up to a possible relative sign).

  As a further simplification, we will neglect the mixing
between CP-even and CP-odd Higgs bosons due to these phases.
While this mixing can be significant in some regions of the 
MSSM parameter space, especially for large values of $\tan\beta$, $|\mu|$ 
and $|A_t|$, and small $M_{H^+}$~\cite{cphiggs}, we have checked
that the mixing (induced by chargino and neutralino loop corrections) 
is small  ($\lesssim 3\%$) 
for the parameters considered here, where $\tan\beta$ takes only 
moderate values and 
the only relevant phase is the one associated with the gaugino sector.   
We also note that in~\cite{Gomez:2004ek} the effect of Higgs mixing 
on the neutralino relic density was 
found to be small, even in the large $\tan\beta$ regime, where the 
Higgs boson mixing is much larger.
The supersymmetric corrections to the bottom mass~\cite{bottomm} 
are also suppressed in the region of parameter space considered
here, and hence all relevant CP violating effects are associated with
the tree-level effect on the neutralino masses and couplings.

\subsection{Relic Density}

We compute the relic abundance of neutralinos by numerically solving the 
Boltzmann equation,
\bea
\frac{dn}{dt} =-3Hn-\langle\sigma_{eff}v\rangle\left(n^2-n_{eq}^2\right),
\label{Eq:Boltzmann}
\eea
for the number density of the supersymmetric particles $n=\sum_{i=1}^N n_i$.
Due to conservation of $R$ parity, the present value of $n$ is equal to the number 
density of the lightest neutralino $n_1$.  In Eq.~(\ref{Eq:Boltzmann})
$H=100h$ km/sec/Mpc, $n_{eq}$ is the value of $n$ at thermal equilibrium, and 
\bea
\langle\sigma_{eff}v\rangle (x) =
\frac{\int_2^\infty K_1\left({ax}\right)
      \sum_{i,j=1}^{N}\lambda(a^2,b_i^2,b_j^2)g_ig_j\sigma_{ij}(a)da}
     {\frac{4}{x}\left( \sum_{i=1}^{N}K_2\left({b_ix}\right) b_i^2 g_i\right)^2} 
\label{Eq:sigv}
\eea
is the thermally averaged annihilation cross section. This quantity is 
a function of
$x = m_{1}/T$, and is given in terms of the
individual annihilation cross sections $\sigma_{ij}(a)$ of the processes 
$ij\to $ SM and/or Higgs particles.  The energy and mass fractions 
$a=\sqrt{s}/m_1$ and $b_i=m_i/m_1$ also enter via $\lambda(a^2,b_i^2,b_j^2) 
= a^4+b_i^4+b_j^4-2(a^2 b_i^2+a^2 b_j^2+ b_i^2 b_j^2)$.
In Eq.(\ref{Eq:sigv}) $g_i$ is the number of degrees of freedom of the 
$i$th supersymmetric partner, and $K_l$ is the modified Bessel function 
of the second kind of order $l$.  The mass of the lightest neutralino is 
denoted by $m_1$.  

In our calculation all relevant annihilation and coannihilation processes are 
included as described in Ref.\cite{Baer:2002fv}.  Besides neutralino 
self-annihilations, coannihilations of the lightest neutralino with the 
lightest stop and the lighter chargino, and annihilations of the lightest 
stop and chargino effect significantly our numerical results.  
The complex phases enter our relic density calculation directly through 
the couplings and indirectly through the masses of 
the neutralinos and charginos.
After diagonalization of the gaugino and sfermion complex mass matrices, we 
calculate the annihilation cross sections with complex couplings.  In doing 
this, we follow techniques used in 
Refs.\cite{Gondolo:2004sc,Katsanevas:1997fb}.

\begin{figure*}[htb]
  \includegraphics[height=8.5cm]{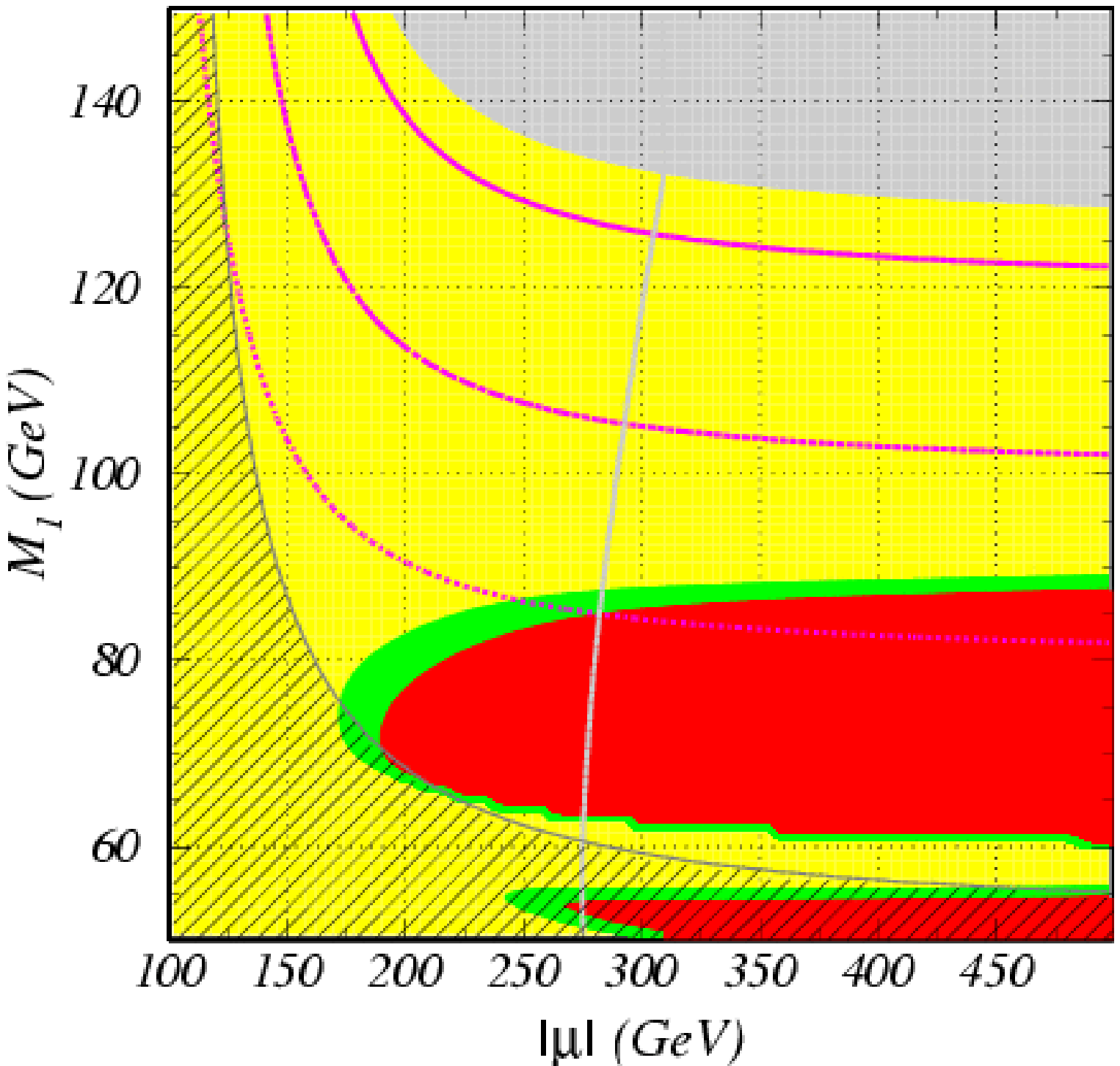}
  \hspace{0.5cm}
  \includegraphics[height=8.5cm]{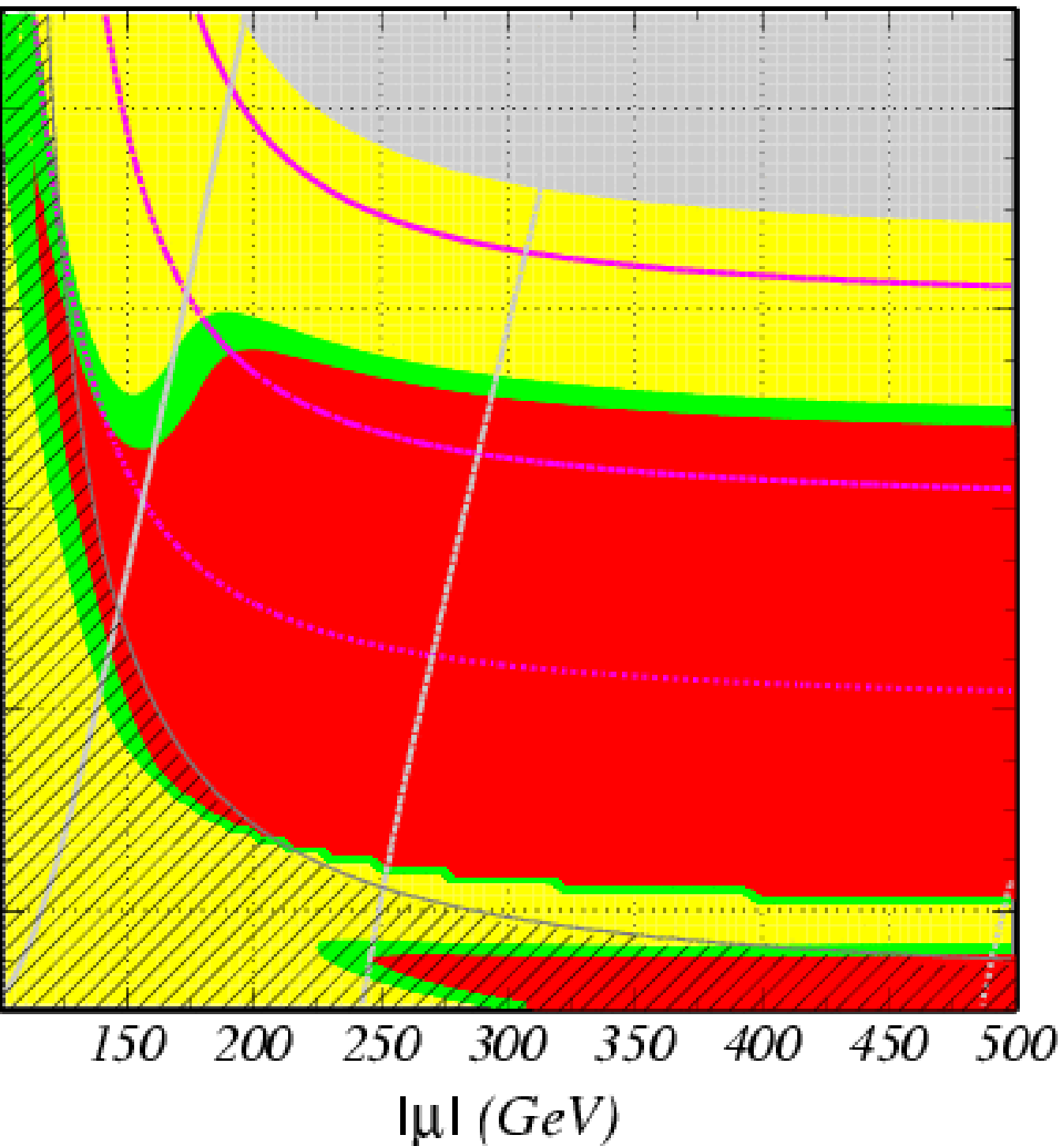}
  \includegraphics[height=1.9cm]{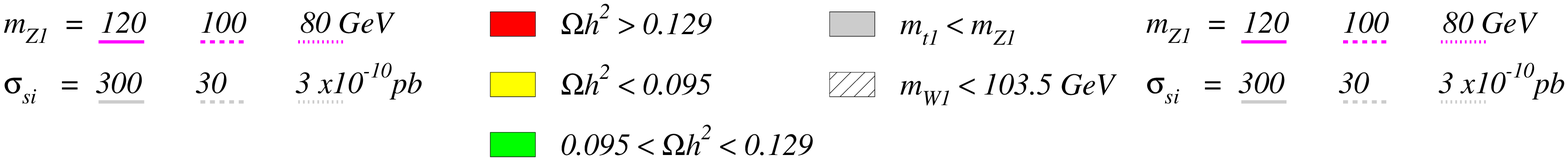}
  \caption{Neutralino relic density for $M_A = 200$~GeV~(left)
and $M_A = 1000$~GeV~(right), and $Arg(\mu)=0$.}
  \label{fig:dm1}
\end{figure*}

\begin{figure*}[htb]
  \includegraphics[height=8.5cm]{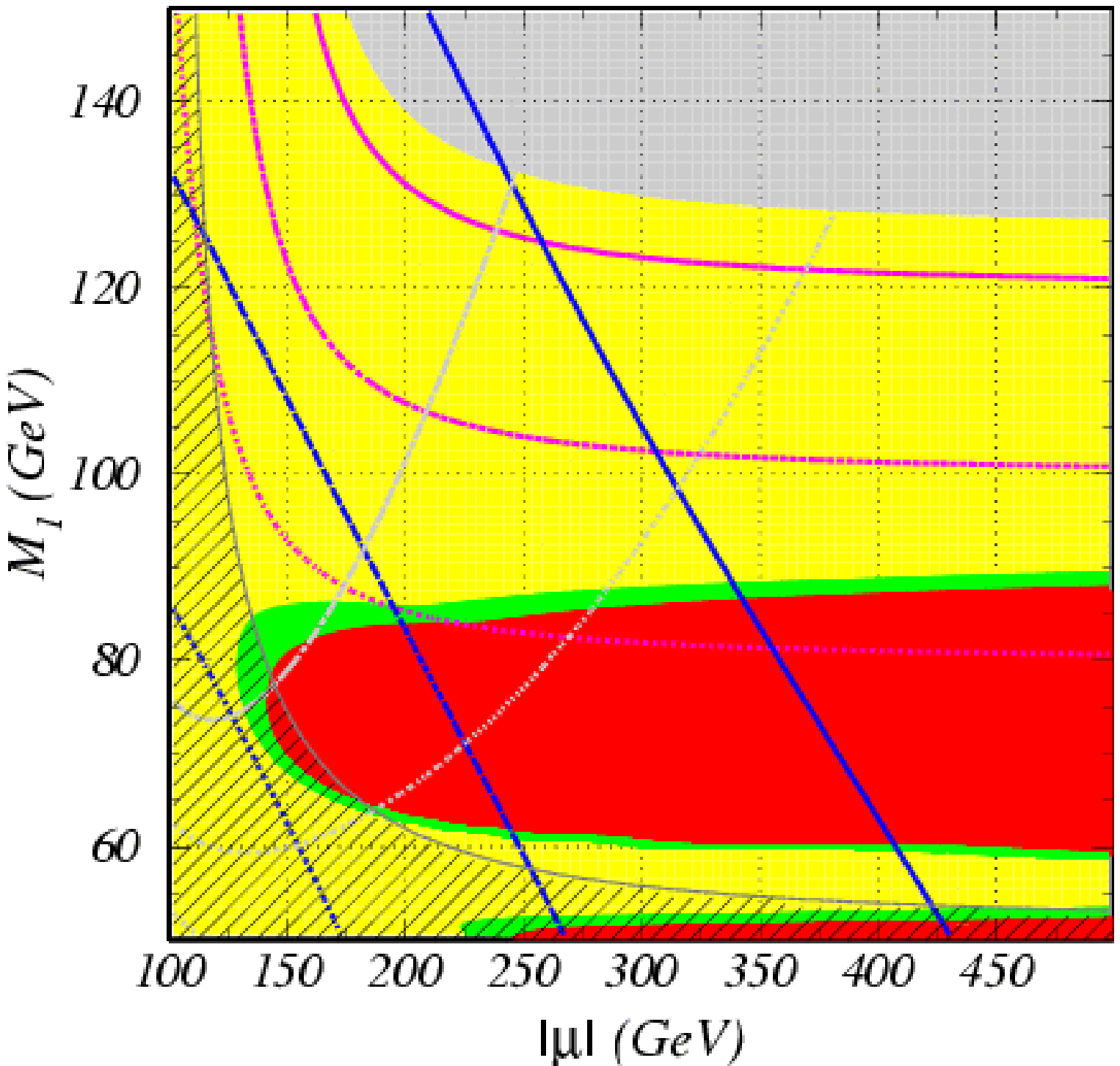}
  \hspace{0.5cm} 
  \includegraphics[height=8.5cm]{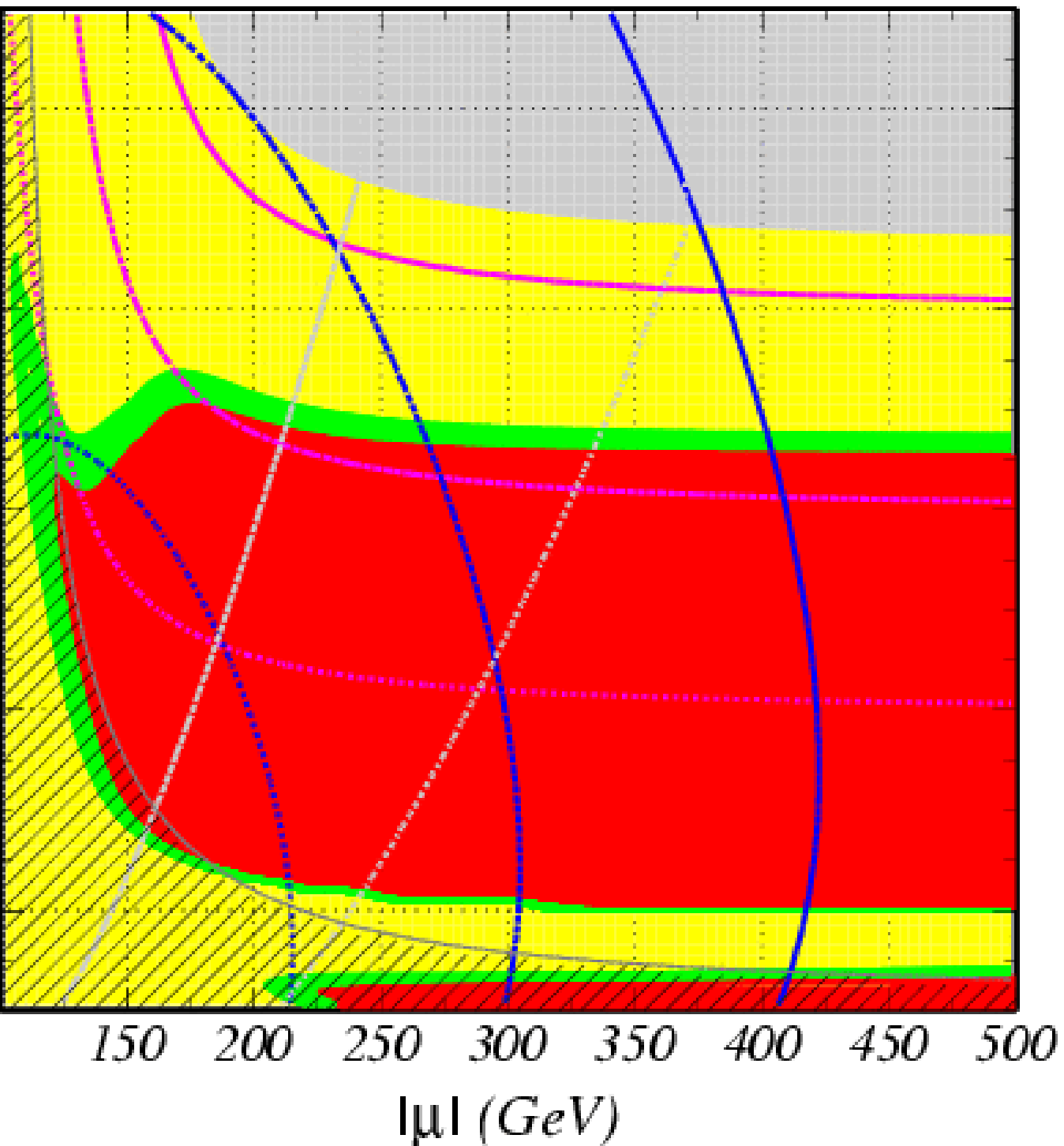}
  \includegraphics[height=1.9cm]{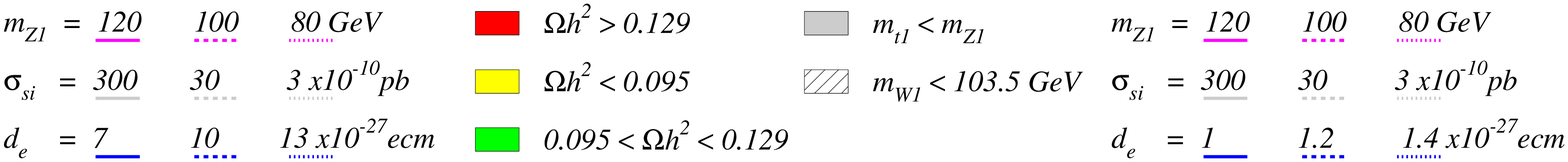}
  \caption{Neutralino relic density for $M_A = 200$~GeV~(left)
and $M_A = 1000$~GeV~(right), and $Arg(\mu)=\pi/2$.}
  \label{fig:dm2}
\end{figure*}

\begin{figure*}[htb]
  \includegraphics[height=8.5cm]{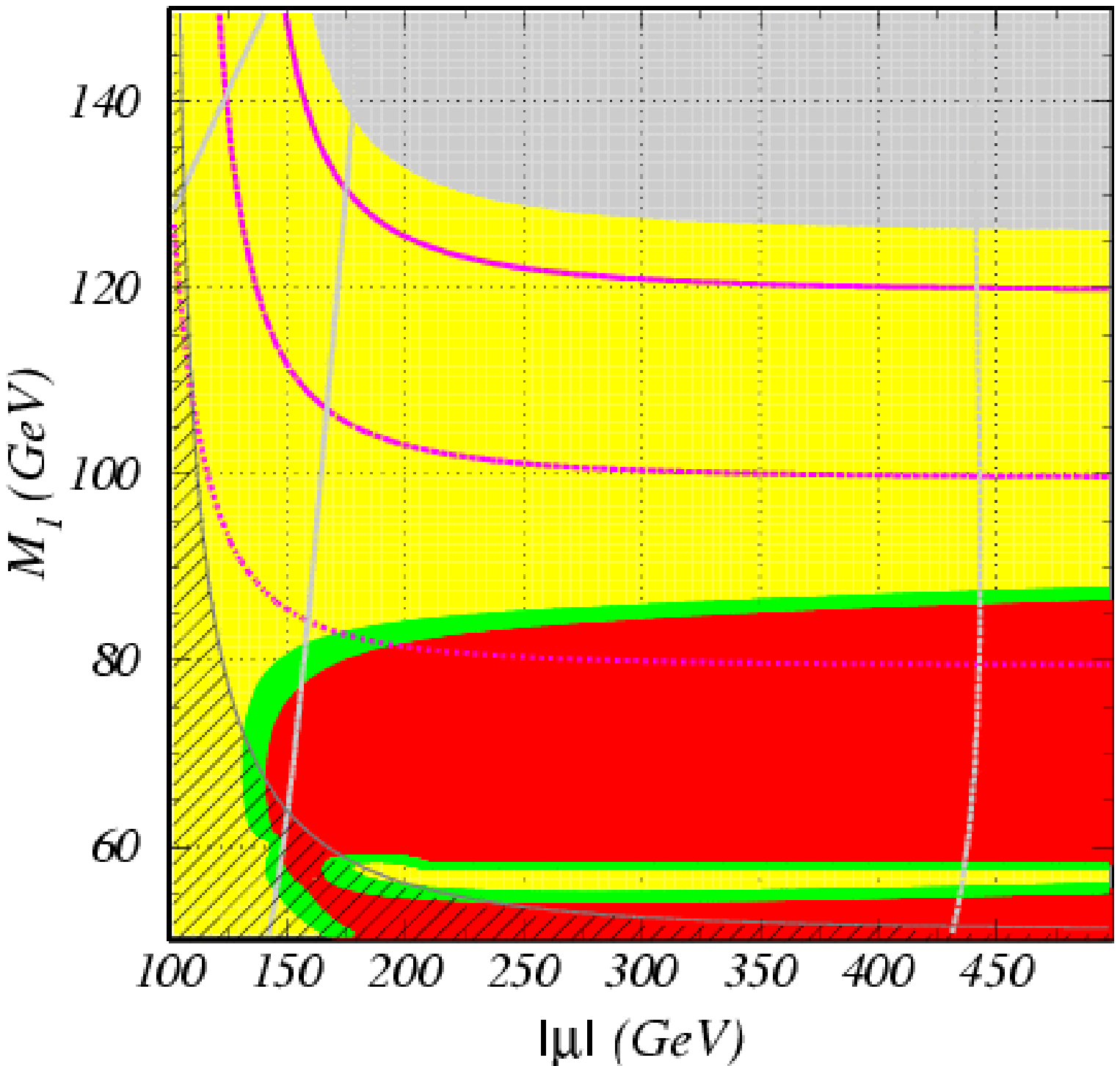}
  \hspace{0.5cm}
  \includegraphics[height=8.5cm]{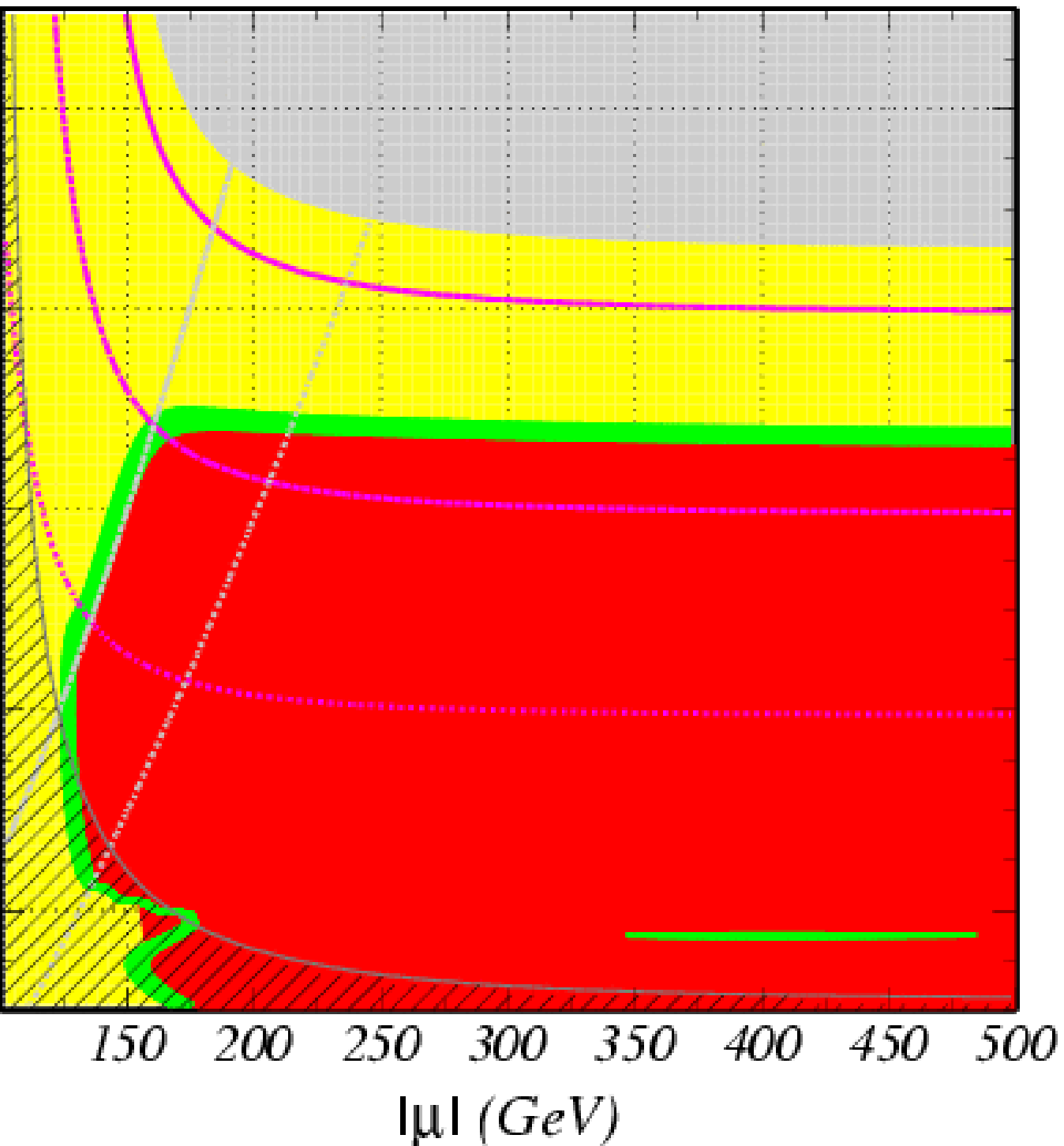}
  \includegraphics[height=1.9cm]{legend0.eps}
  \caption{Neutralino relic density for $M_A = 200$~GeV~(left)
and $M_A = 1000$~GeV~(right), and $Arg(\mu)=\pi$.}
  \label{fig:dm3}
\end{figure*}

  Figures~\ref{fig:dm1}-\ref{fig:dm3} show the dependence of the
neutralino relic density on $|\mu|$ and $M_1$ for $\tan\beta = 7$, $M_A = 
200$~GeV (left) and $M_A = 1000$~GeV (right), and three values of the 
$\mu$ phase: $Arg({\mu}) = 0,\;\pi/2,\;\pi$.  
Values of the phase equal to $0$ or $\pi$ are representative of what 
happens for small phases, like the ones consistent with the generation 
of the baryon asymmetry when $|\mu| \simeq M_2$ and $M_A\lesssim 300$~GeV
where there is a resonance in the amount of baryon number 
produced~\cite{Carena:2002ss}. 
On the other hand, large values of the phase, close to $\pi/2$, 
tend to be necessary to generate the baryon asymmetry outside of the 
resonant region, particularly for large values of $M_A$, for which the EDM 
constraints become less severe.

The green (medium gray) bands in Figures~\ref{fig:dm1}-\ref{fig:dm3}
show the region of parameter space where the neutralino relic 
density is consistent with the $95\%$~CL limits set by WMAP data. 
The regions in which the relic density is above 
the experimental bound and excluded by more than two standard 
deviations are indicated by the red (dark gray) areas.  
The yellow (light gray) areas show the regions
of parameter space in which the neutralino relic density is less 
than the WMAP value.  An additional source of dark matter,
unrelated to the neutralino relic density, would be needed in
these regions.  Finally, in the (medium-light) gray region at the upper right 
the lightest stop becomes the LSP, while in the hatched area
at the lower left corner the mass of the lightest chargino 
is lower than is allowed by LEP data~\cite{LEPSUSYW1}. 

  These figures are qualitatively similar, but do show
some differences due to the change in the phase of $\mu$.
Before discussing the effect of the phase, we will examine
the general features of Figures~\ref{fig:dm1}-\ref{fig:dm3}.
For $M_A = 1000$~GeV and for all three phase values, the region where the 
relic density is too high consists of a wide band in which the lightest 
neutralino has mass between about $60$~and~$105$~GeV and is predominantly Bino.
Above this band, the mass difference between the neutralino 
LSP and the light stop is less than about $20$~GeV, and stop-neutralino 
coannihilation as well as stop-stop annihilation are very efficient at 
reducing the neutralino abundance.  
For $M_A = 200$~GeV, instead, the contribution to neutralino annihilation 
from s-channel exchange of heavy CP-even and CP-odd Higgs bosons 
is enhanced by a resonance around $m_{\tilde{Z}_1} \simeq 100$~GeV.  
This restricts the band in which the relic density is too high to 
the region where the lightest neutralino has mass between about 
$60$~and~$85$~GeV, and is also mostly Bino.  
For both values of $M_A$,
there is an area below the disallowed band in which 
the neutralino mass lies in the range $40$-$60$~GeV, and the neutralino 
annihilation cross-section is enhanced by resonances from 
s-channel $h^0$ and $Z^0$ exchanges.

The relic density is also quite low for smaller values of $|\mu|$.  
In these regions, the neutralino LSP acquires a significant Higgsino component 
allowing it to couple more strongly to the Higgs bosons and the $Z^0$. 
For $M_A = 1000$~GeV, this is particularly important 
in the region near  $(|\mu|, M_1) = (175,110)$~GeV
where the neutralino mass becomes large enough that annihilation into 
pairs of gauge bosons through s-channel Higgs and $Z^0$ exchange
and t-channel neutralino and chargino exchange is allowed,
and is the reason for the dip in the relic density near this point.  
Since the corresponding couplings to the gauge bosons depend on the 
Higgsino content of the neutralino,
these decay channels turn off as $|\mu|$ increases. 
For higher $M_1$ values, the lightest neutralino and chargino masses 
are also close enough that chargino-neutralino coannihilation and 
chargino-chargino annihilation substantially increase the effective cross section.

  In Figures~(\ref{fig:dm1}-\ref{fig:dm3}), we have taken $M_2 = (g_2^2/g_1^2)M_1$,
as suggested by universality.  Because of this, smaller values of $M_1$ and 
$\mu$ are excluded by the lower bound on the chargino mass from LEP 
data~\cite{LEPSUSYW1}, as indicated by the hatched regions in the figures.  
This constraint becomes much less severe for larger values of the 
ratio $M_2/M_1$.  We also find that increasing this ratio of gaugino masses
(with $M_1$ held fixed) has only a very small effect 
on the neutralino relic density.

\subsection{Effects of CP Violating Phases}

  For the parameters considered in the previous section, 
relevant for EWBG within the MSSM, CP violating phases modify the 
values of the neutralino relic density but have only a mild
effect on the general qualitative features of the 
allowed parameter space.  This is somewhat misleading, however, 
since the value of the relic density at a given point 
in the $|\mu|\!-\!M_1$ plane can vary markedly with $Arg(\mu)$.

\begin{figure*}[htb]
\centerline{
        \includegraphics[width=0.5\textwidth]{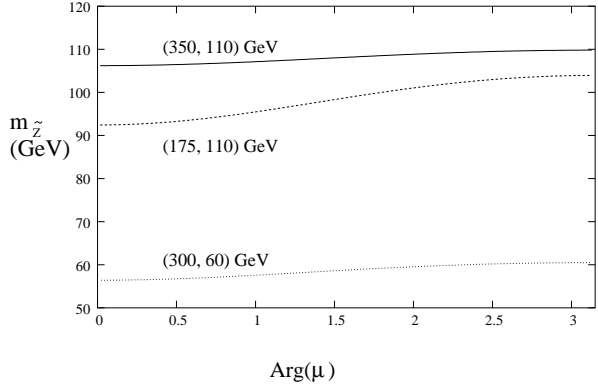}}
        \caption{Mass of the lightest neutralino as a
function $Arg(\mu)$ for $\tan\beta = 7$
and three sample values of $(|\mu|, M_1)$.}
\label{nmass}        
\end{figure*}

  The most important effect of varying $Arg(\mu)$ is to 
shift the mass of the neutralino LSP.  The dependence
of the lightest neutralino mass on this phase is shown
in Figure~\ref{nmass} for $\tan\beta = 7$ and
three sample values of $(|\mu|,M_1)$: 
$(|\mu|,M_1) = (350,110)$~GeV, $(300,60)$~GeV,
and $(175,110)$~GeV. For $M_A = 1000$~GeV, these three points
are representative of the regions where the annihilation cross
section is dominated by stop-neutralino coannihilation~($(350,110)$~GeV), 
Higgs boson s-channel annihilation~($(300,60)$~GeV), and annihilation 
into pairs of gauge bosons~($(175,110)$~GeV). 
In all three cases, the neutralino mass increases
with $Arg(\mu)$, by about 3\%, 7\%, and 11\%, respectively. 
Such a mass shift can significantly modify the relic density
at a single point where neutralino annihilation is enhanced
by a resonance or coannihilation with another species.
The effect on the net distribution of relic densities, on the other hand,
is fairly small; shifting the phase tends to translate this distribution
down and to the left in the $|\mu|$-$M_1$ plane.

  The neutralino-Higgs couplings are also quite sensitive to $Arg(\mu)$.  
The couplings of the Higgs bosons to a pair of 
neutralinos are given in~\cite{Gunion:1989we}, and have the form
\bea
{\tilde{Z}_1}\tilde{Z}_1\,h^0/H^0 \:\:&\sim&\:\: -i(F\:P_L + F^*P_R)\nonumber\\
{\tilde{Z}_1}\tilde{Z}_1\,A^0 \:\:&\sim&\:\: -i(G\:P_L - G^*P_R)
\label{Eq:nnhiggs}
\eea
where $P_{L,R} = (1 \mp \gamma_5)/2$ are the usual chiral projectors.
Using these vertices, the spin-summed and squared
matrix elements for $\tilde{Z}_1\tilde{Z}_1 \to \bar{f}f$
annihilation via s-channel Higgs exchange are proportional to
\be
\left|\mathcal{M}\right|^2 \propto\left\{
\begin{array}{ll} 
Re(F)^2\,(s-4\,m_{\tilde{Z}_1}^2)\;+\; Im(F)^2\,s\;;\:&\hspace{1cm}h^0,~H^0\\
Re(G)^2\,s\;+\; Im(G)^2\,(s-4\,m_{\tilde{Z}_1}^2)\;;\:&\hspace{1cm}A^0
\end{array}\right.
\label{Eq:matel}
\ee
In calculating the thermal average, one integrates these 
matrix elements over $s$ through the range 
$[4\,m_{\tnt}^2,\infty)$ with a Boltzmann factor, Eq.~(\ref{Eq:sigv}).  
The Boltzmann suppression is strong for a cold relic, 
so the integral is dominated by the region $s\sim 4\,m_{\tnt}^2$.  
In particular, this means that the terms in Eq.~(\ref{Eq:matel}) 
proportional to $s$ have the potential to give a much 
larger contribution to the thermal average than 
those proportional to $(s-4\,m_{\tnt}^2)$.

\begin{figure*}[htb]
\centerline{
        \includegraphics[width=0.5\textwidth]{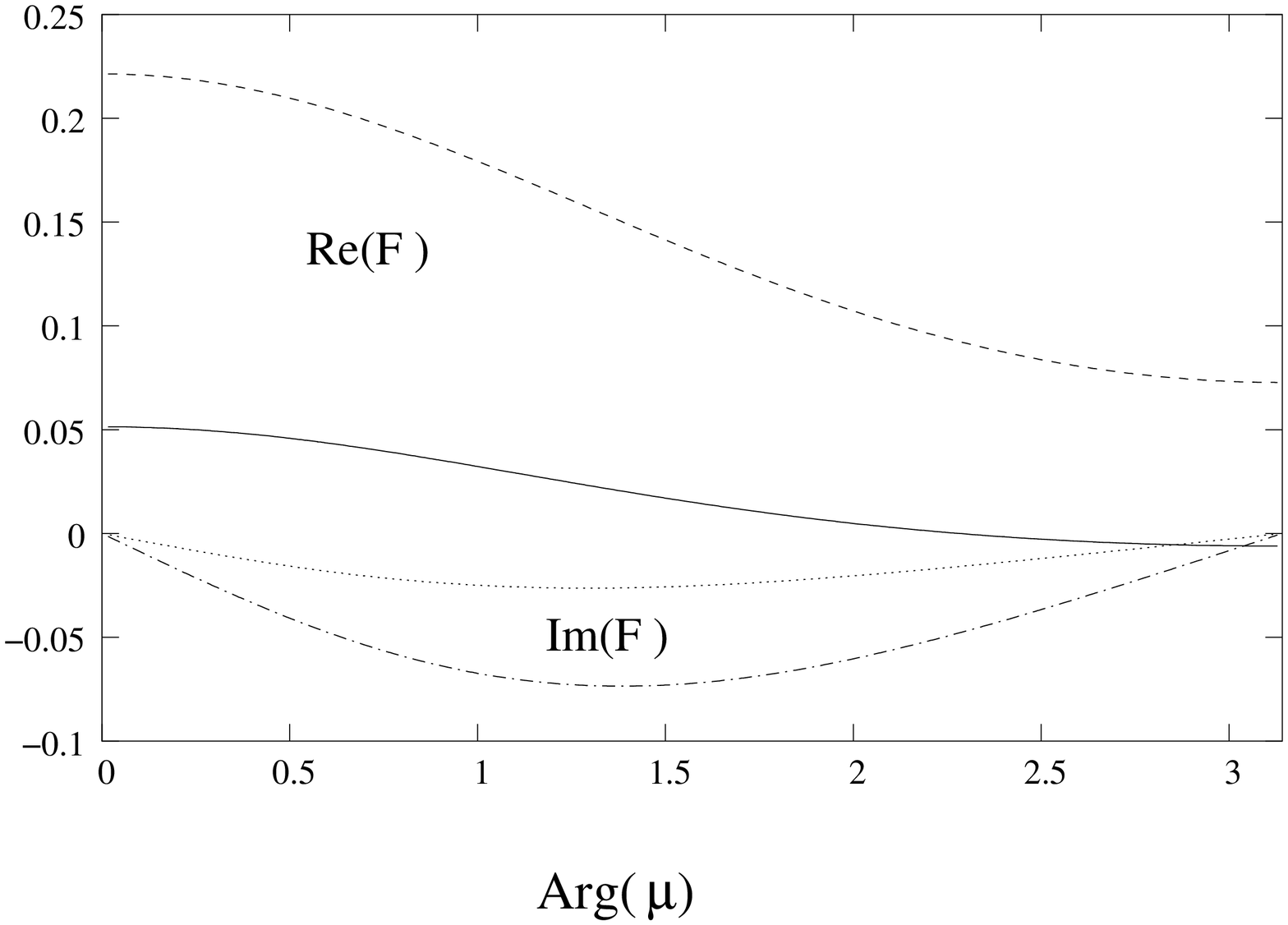}}
        \caption{Variation of the real and imaginary parts
of the $\tilde{Z}_1\tilde{Z}_1h^0$ coupling with $Arg(\mu)$
for $(|\mu|,M_1) = (300,60)$~GeV~(solid and dotted), and
$(|\mu|, M_1) = (175,110)$~GeV~(dashed and dash-dotted).}
\label{Fig:hcoup}        
\end{figure*}

\begin{figure*}[hbt!]
\begin{minipage}[t]{0.47\textwidth}
        \includegraphics[width = \textwidth]{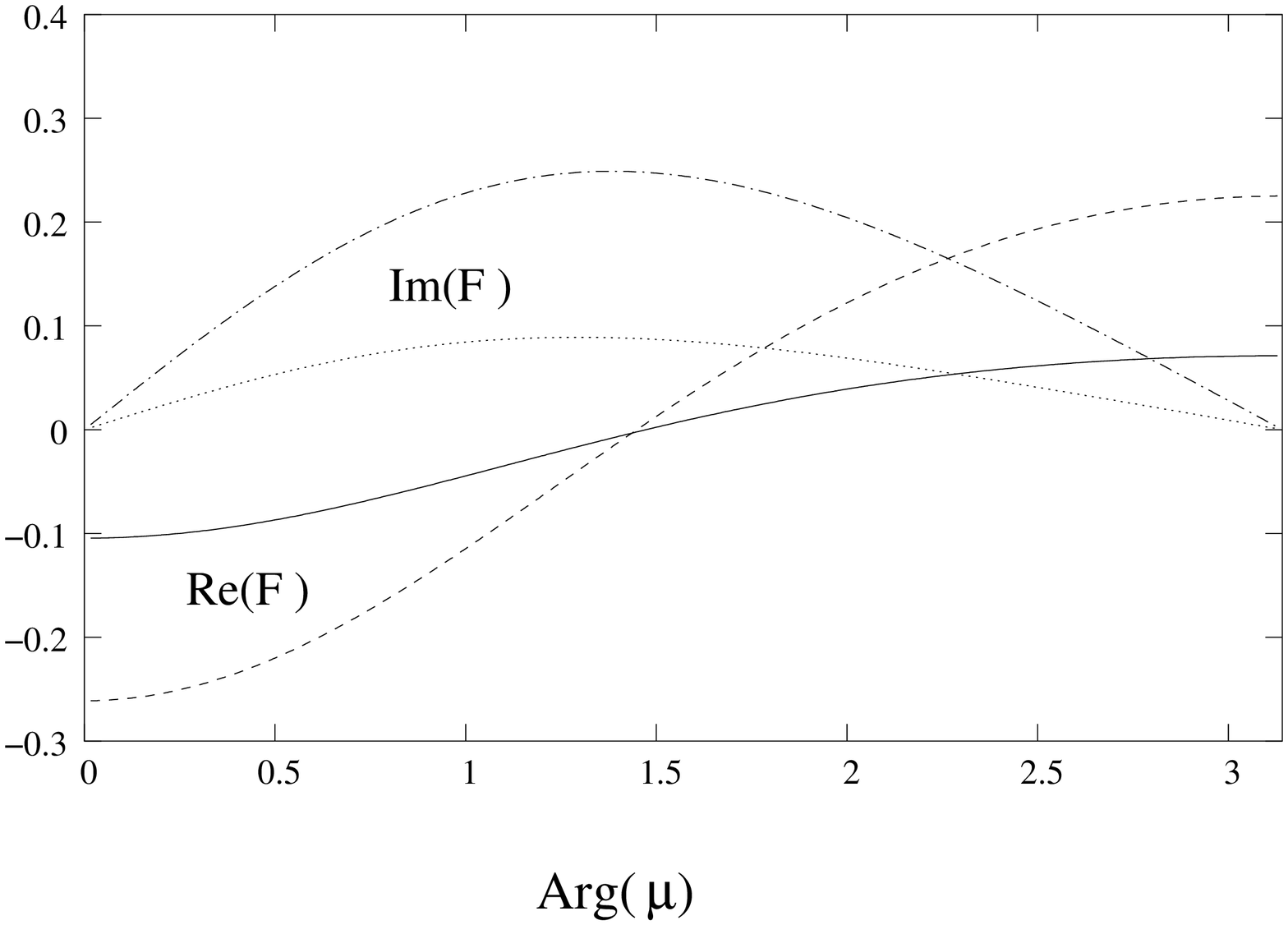}
\end{minipage}
\phantom{aa}
\begin{minipage}[t]{0.47\textwidth}
        \includegraphics[width = \textwidth]{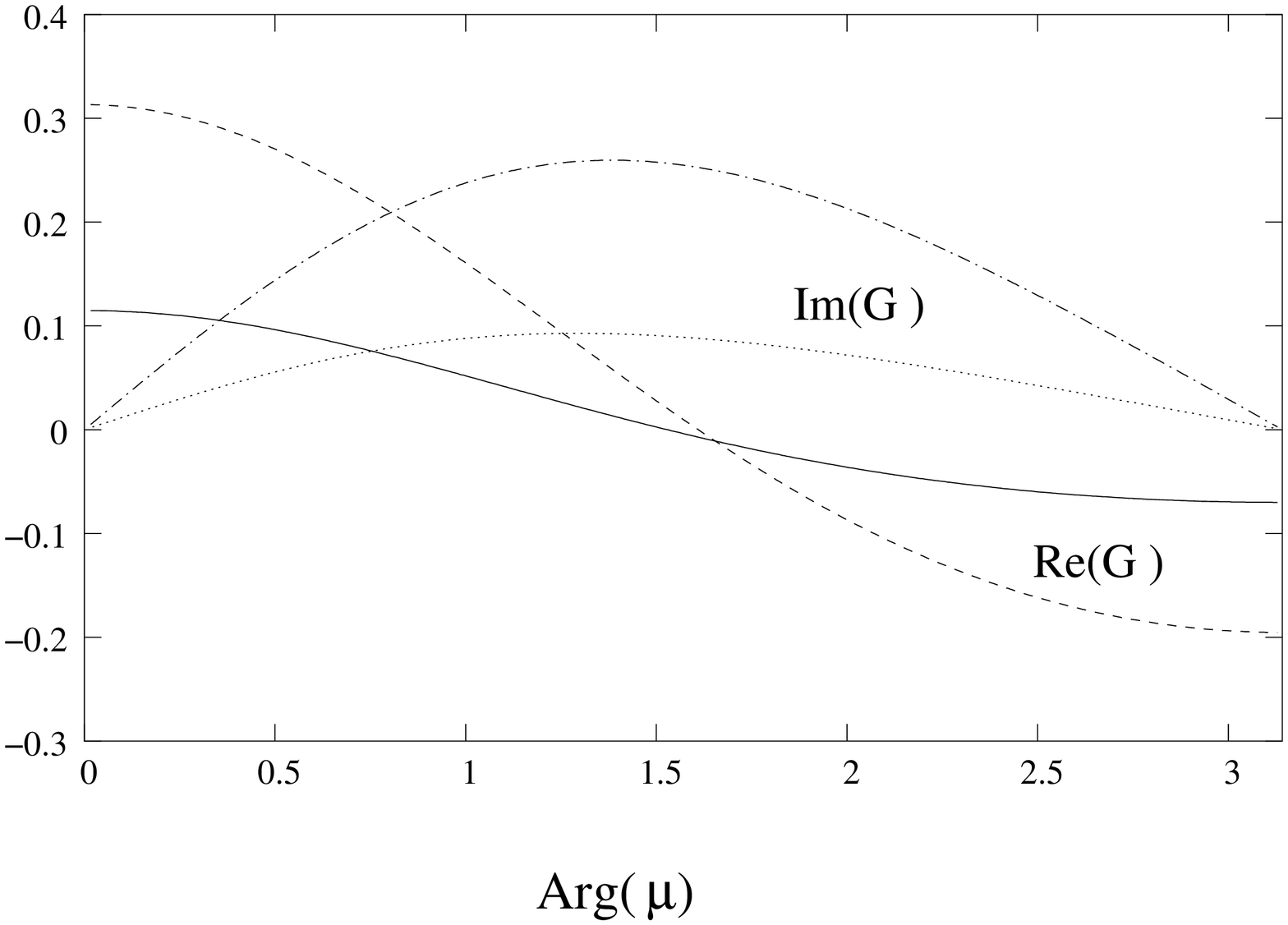}
\end{minipage}
\caption{Variation of the real and imaginary parts
of the $\tilde{Z}_1\tilde{Z}_1H^0$~(left) 
and $\tilde{Z}_1\tilde{Z}_1A^0$~(right) couplings with $Arg(\mu)$
for $(|\mu|,M_1) = (300,60)$~GeV~(solid and dotted), and
$(|\mu|, M_1) = (175,110)$~GeV~(dashed and dash-dotted).}
\label{Fig:Hcoup}
\end{figure*}

  The dependence of the ${\tilde{Z}_1}\tilde{Z}_1\,h^0$ coupling on $Arg(\mu)$ for 
$M_A=1000$~GeV, and $(|\mu|, M_1) = (300,60)$~GeV and $(|\mu|, M_1) 
= (175,110)$~GeV is shown in Figure~\ref{Fig:hcoup}.
Both the real and imaginary parts of the couplings are larger in
the $(|\mu|, M_1) = (175,110)$~GeV case since for these values of the
 parameters, the neutralino
LSP has a much larger Higgsino component than for
$(|\mu|, M_1) = (300,60)$~GeV,
when the neutralino is mostly Bino.  
The couplings for $(|\mu|, M_1) = (350,110)$ GeV, where the LSP is also mostly
Bino, are very similar to those for $(|\mu|, M_1) = (300,60)$~GeV.  
Setting $M_A = 200$~GeV has only a small effect on these couplings.
For both points shown in Figure~\ref{Fig:Hcoup}, 
the imaginary part of the coupling
vanishes when $\mu$ is real, and is largest when $\mu$ is pure
imaginary, $Arg(\mu) = \pi/2$.  The real part of the coupling
also tends to decrease with $Arg(\mu)$ due to an accidental 
cancellation of terms.  
This behavior may be seen by comparing the region 
$M_1\lesssim 60$~GeV in Figures~\ref{fig:dm1}, \ref{fig:dm2}, 
and \ref{fig:dm3}, where s-channel $h^0$ exchange 
tends to be dominant.  The relic density in this region is 
lowest when $Arg(\mu) = \pi/2$, Figure~\ref{fig:dm2},
while in Figure~\ref{fig:dm3}, corresponding to $Arg(\mu) = \pi$,
the contribution from $h^0$ exchange is much smaller 
than for other values of this phase.  

  The couplings of the $H^0$ and $A^0$ bosons to neutralinos are shown in
Figure~\ref{Fig:Hcoup} for $M_A=1000$~GeV, and 
$(|\mu|, M_1) = (300,60)$~GeV and $(|\mu|, M_1) = (175,110)$~GeV.  
As with the $h^0$ coupling, these couplings are nearly unchanged when $M_A= 200$~GeV,
and the couplings for $(|\mu|, M_1) = (350,110)$~GeV are very similar to
those for $(|\mu|, M_1) = (300,60)$~GeV.
The imaginary part of the $H^0$ and $A^0$ couplings vanishes for 
$Arg(\mu) = 0,\;\pi$ and is largest near $Arg(\mu) = \pi/2$, while  
the real parts of these couplings are largest for $Arg(\mu) = 0,\;\pi$
and nearly zero when $Arg(\mu) = \pi/2$.  From Eqs.~(\ref{Eq:nnhiggs},
\ref{Eq:matel}), this implies that the contribution of s-channel $H^0$ 
exchange to neutralino annihilation is largest when $Arg(\mu) = \pi/2$,
and smallest for $Arg(\mu) = 0,\;\pi$, and that the opposite is true
for s-channel $A^0$ exchange.  Interestingly, the sum of the $A^0$ and $H^0$
contributions is nearly independent of the phase.  We expect this to be
the case whenever $M_A^2 \gg M_Z^2$, and the heavy CP-even and CP-odd Higgs states
are nearly degenerate.  The same effect was found in~\cite{Gomez:2004ek}.

  We have also investigated the phase dependence of the 
$\tilde{Z}_1\,t\,\tilde{t}$ coupling which generates the most
important contributions to stop-neutralino coannihilation.
While this coupling does vary somewhat with the phase, the effect
of the phase on the neutralino mass is much more important.
This is because the coannihilation contribution to the relic density
is suppressed by a factor $e^{-(m_{\tilde{t}}-m_{\tilde{Z}_1})/T_f}$,
where $T_f \simeq m_{\tilde{Z}_1}/20$ is the neutralino freeze-out temperature,
making it very sensitive to the neutralino mass.

\section{Direct Detection of Dark Matter \label{direct}}

If space around us is filled with relic neutralinos, then it is plausible to 
try to observe them. Indeed, the search for weakly interacting massive 
particles is in progress via detection of their scattering off nuclei by 
measuring the nuclear recoil.  Since neutralinos are non-relativistic they 
can be directly detected via the recoiling off a nucleus in elastic 
scattering. 
There are several existing and future experiments engaged in this search. 
These include solid state germanium, ionization based detectors such as 
IGEX~\cite{Irastorza:2002vk}, HDMS~\cite{hdms}, CDMS~\cite{CDMS}, 
EDELWEISS~\cite{EDELWEISS} and GENIUS~\cite{GENIUS}.  Solid crystal or 
liquid NaI based scintillator detectors are used for example by 
DAMA~\cite{DAMA} and ZEPLIN~\cite{ZEPLIN1,ZEPLIN2,ZEPLIN3,ZEPLIN4}.  Liquid, 
gas or hybrid xenon based detector is used by experiments as XENON 
\cite{Aprile:2004ey} and UKDMC \cite{Spooner:2001hn}.  Gas target projection 
chambers are utilized in DRIFT~\cite{spooner}, and metastable particle 
detectors in SIMPLE~\cite{simple} and PICASSO~\cite{picasso}. 

\begin{figure*}[htb]
\centerline{\includegraphics[width=0.8\textwidth]{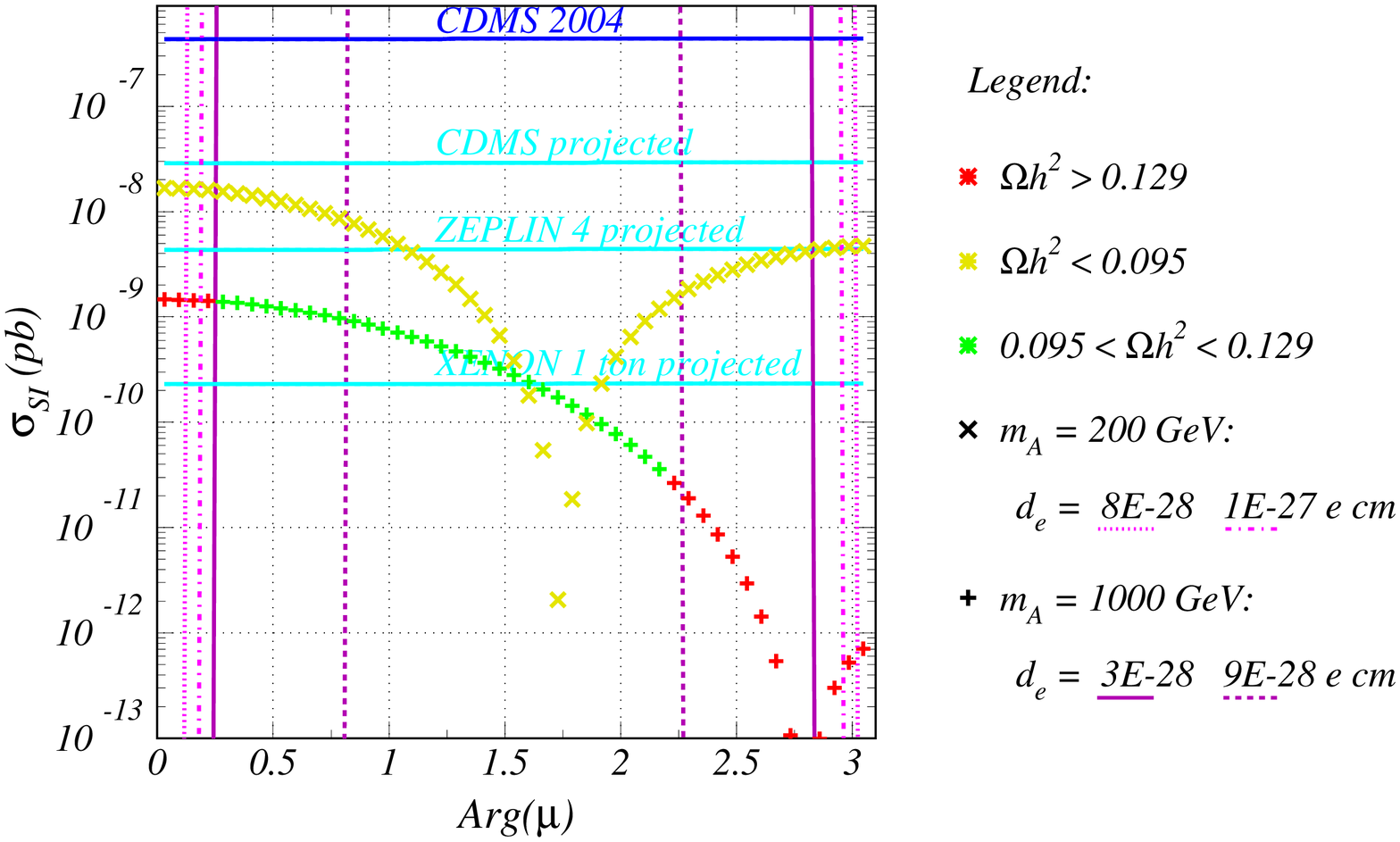}}
\caption{Spin-independent neutralino-proton scattering cross 
section as the function of $Arg(\mu)$, for $|\mu|=$~350~GeV and 
$M_1 = 110$~GeV, and for $m_A$= 200 (1000) GeV for the upper (lower) curve.}
\label{fig:dd1}        
\end{figure*}

The elastic scattering interactions of neutralinos with nuclei can be 
described by the sum of spin independent (${\cal L}^{eff}_{SI}$) and spin 
dependent (${\cal L}^{eff}_{SD}$) Lagrangian terms:
\bea
 {\cal L}^{eff}_{elastic}={\cal L}^{eff}_{SI}+{\cal L}^{eff}_{SD} .
\eea
For heavy nuclei the spin independent (SI) cross section, being proportional 
to the squared mass of the target nucleus, is highly enhanced compared to 
the spin dependent one.  For the case of a target containing the isotope 
$^{127}$I, for example, the enhancement factor is more than $10^4$.  For 
this reason the experimental limits on the spin independent 
neutralino-nucleon cross sections are considerably stronger.

\begin{figure*}[htb]
\centerline{
        \includegraphics[width=0.8\textwidth]{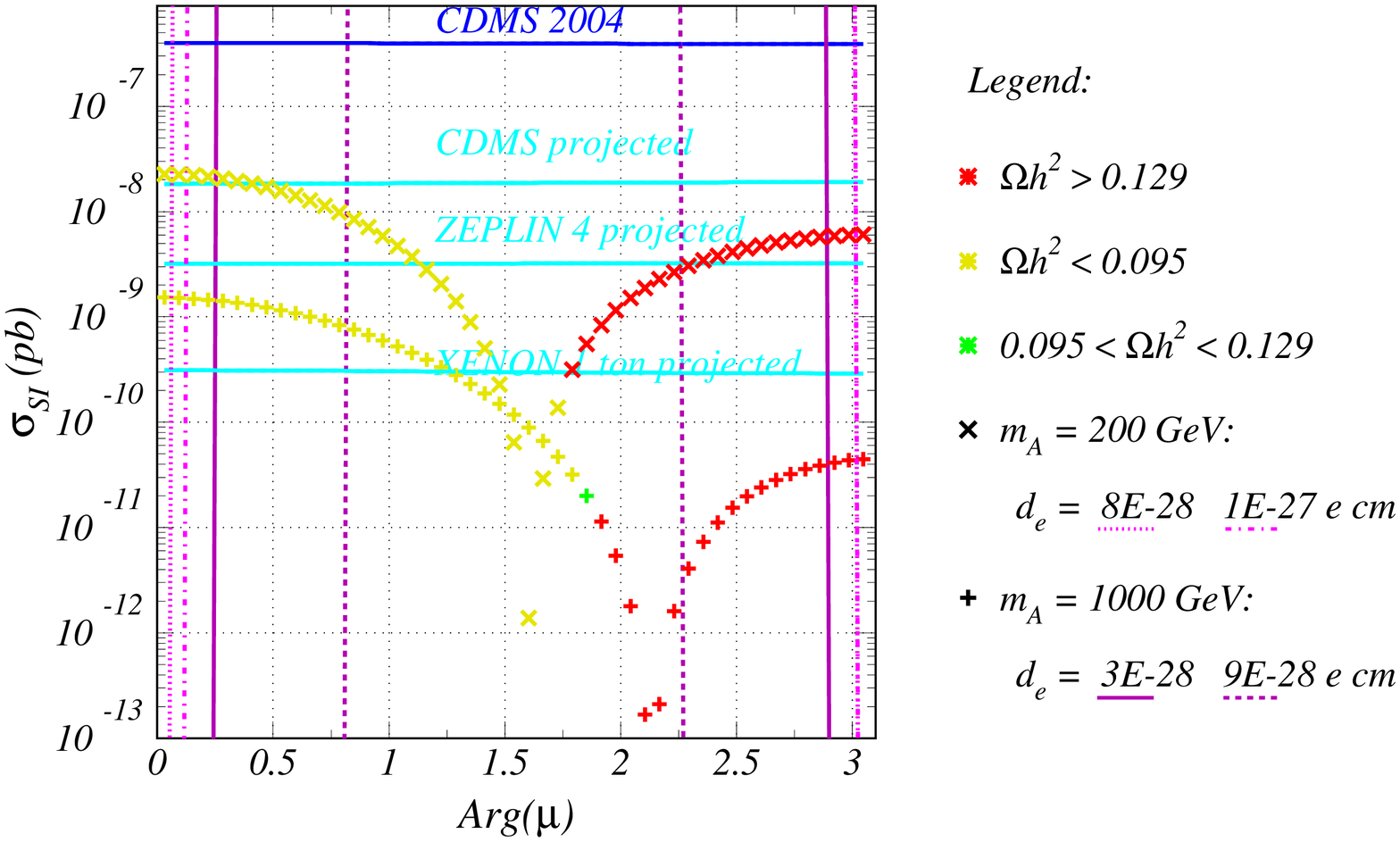}}
        \caption{Same as Figure~\ref{fig:dd1}, but 
for $|\mu| = 300$~GeV and 
$M_1=60$~GeV.}
\label{fig:dd2}        
\end{figure*}

In what follows, we will focus on the spin independent interactions of 
neutralinos with nuclei.  At the parton level, these are mediated by 
t-channel Higgs and s-channel squark exchanges.  (Here, we only consider the, 
so called, scalar contribution and neglect the higher order tensor 
contribution originating from loop diagrams.)  The differential scattering 
rate of a neutralino off a nucleus $X_Z^A$ with mass $m_X$ takes the 
form~\cite{Baer:2003jb}:
\begin{equation}
 \frac{d\sigma_{SI}}{d|\vec{q}|^2}=\frac{1}{\pi v^2}[Z f_p +(A-Z) f_n]^2 
 F^2 (Q_r),
\end{equation}                                 
where $\vec{q}=\frac{m_X m_{\widetilde Z_1}}{m_X+m_{\widetilde Z_1}}\vec{v}$ 
is the three-momentum transfer, $Q_r=\frac{|\vec{q}|^2}{2m_N}$, and 
$F^2(Q_r)$ is the scalar nuclear form factor, $\vec{v}$ is the velocity of 
the incident neutralino and $f_p$ and $f_n$ are effective neutralino 
couplings to protons and neutrons respectively.  The same formalism was used 
in Ref.~\cite{Baer:2003jb} to calculate neutralino-nucleon cross sections, 
and the reader is directed there for further details.  Since modern 
experiments express their limits in terms of the neutralino-proton cross 
section, we calculate and plot this quantity in this work.

To study the dependence of the neutralino-proton cross section on complex 
phases of various supersymmetric parameters, we select a point in the 
examined parameter region where constraints from EWBG, the electron EDM and 
WMAP are simultaneously satisfied.  Specifically, we 
examine values of $M_A = 200,\:1000$~GeV and the same  Higgsino and 
neutralino mass parameters
chosen before, namely $(|\mu|,M_1) = (175,110)$~GeV, 
(350,110)~GeV and (300,60)~GeV. 
As emphasized before, for $M_A = 1000$~GeV these
points correspond to regions in which the annihilation cross section
is dominated by weak processes, coannihilation with
the light stop, and s-channel Higgs exchange, respectively. 

\begin{figure*}[htb]
\centerline{
        \includegraphics[width=0.8\textwidth]{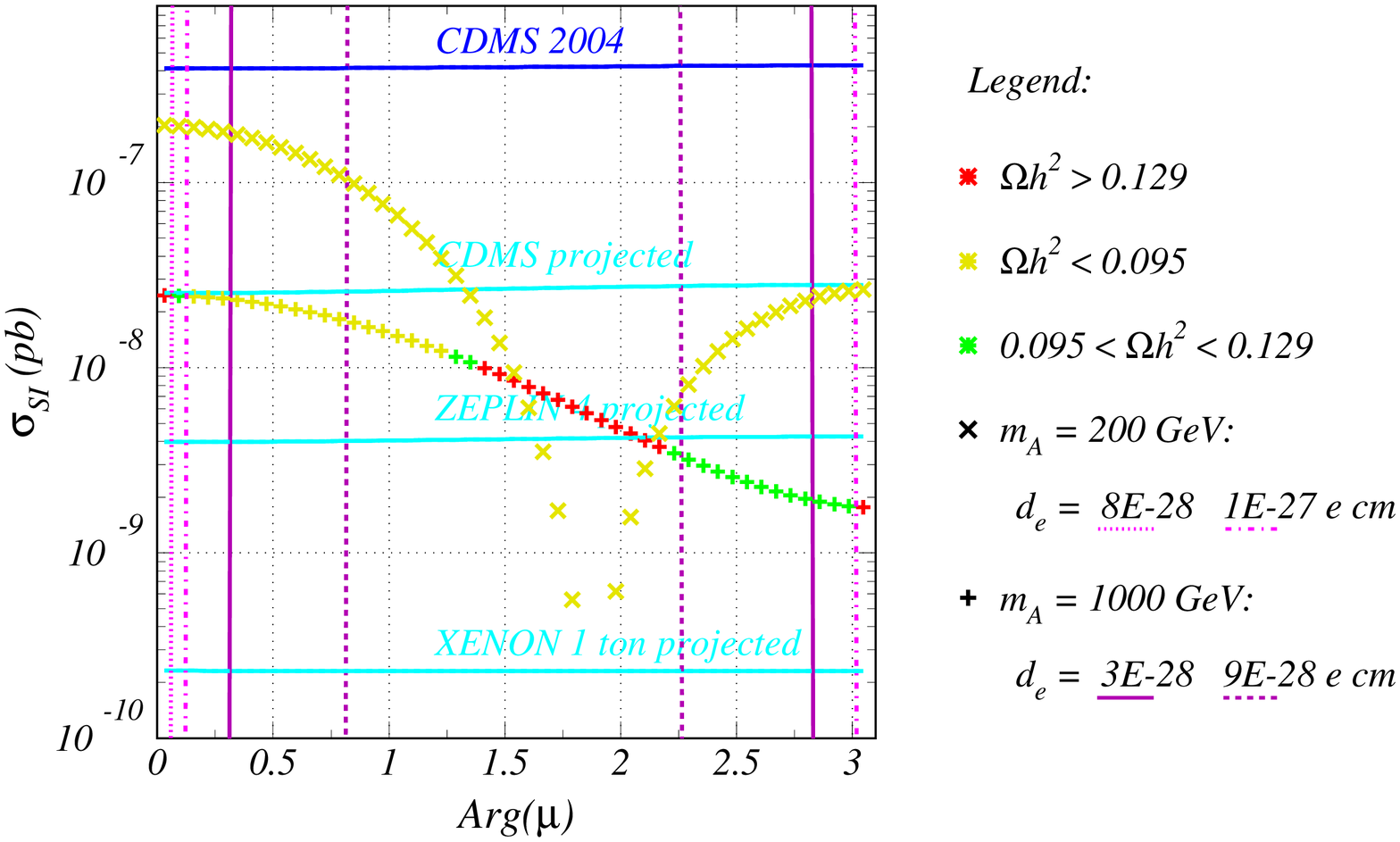}}
        \caption{Same as Figure~\ref{fig:dd1}, but 
for $|\mu| = 175$~GeV and
$M_1 = 110$~GeV.  }
\label{fig:dd3}        
\end{figure*}

Figures~\ref{fig:dd1}--\ref{fig:dd3} show the neutralino-proton cross
section versus the phase of $\mu$ for the selected parameter space points. 
The most striking feature of these plots is that the cross section is 
suppressed for non-vanishing phases and, except for $(|\mu|,M_1) = 
(175,110)$~GeV with $M_A = 1000$~GeV, nearly vanishes for a 
given value of $Arg(\mu)$. 
This behavior follows from the phase dependence of the Higgs-neutralino couplings.
In our case, t-channel $h^0$ and $H^0$ exchange diagrams generate the 
most important contributions to the spin-independent neutralino-nucleon 
elastic scattering cross section.  (We checked that the only relatively light squark, 
the lightest stop, contributes only at the percent level via its s-channel diagram.)
Furthermore, these contributions depend only on the real (scalar) part of the 
Higgs-neutralino couplings~\cite{Choi:2000kh,Nihei:2004bc};
$Re(F)$ in the notation of Eq.~(\ref{Eq:nnhiggs}).  
The large suppression of the cross section for particular values of 
$Arg(\mu)$ is due to zeroes of $Re(F)$.

   Consider first the $M_A = 1000$~GeV lines in Figures~\ref{fig:dd1}--\ref{fig:dd3}.  
For these, $M_A\simeq M_H \gg m_h$, so the contribution of the heavier scalar 
Higgs is suppressed relative to the lighter state, and the neutralino-proton 
scattering is dominated by t-channel $h^0$ exchange.  
Comparing the real part of the $h^0$-neutralino coupling for
$(|\mu|,M_1) = (300,60)$~GeV shown in Figure~\ref{Fig:hcoup}
to the plot of $\sigma_{SI}$ in Figure~\ref{fig:dd2} for $M_A = 1000$~GeV,
we see that the minimum in $\sigma_{SI}$ nearly coincides with the 
zero of the coupling.  
The minimum (not a zero value) in Figure \ref{fig:dd2} does not exactly
coincide with the zero of the coupling, but is shifted closer to 
$Arg(\mu) = \pi/2$ because the zero value of the real part of the 
$H^0$-neutralino coupling occurs close to $Arg(\mu) = \pi/2$, 
as shown in Figure~\ref{Fig:Hcoup}.\footnote{
If the heavy Higgs state is decoupled completely, 
we find that the minimum of the scattering cross section coincides 
exactly with the zero of the $h^0$-neutralino coupling.}  
When $(|\mu|, M_1)=(175,110)$~GeV, the coupling of the lightest Higgs to the 
lightest neutralino has no zero, and $\sigma_{SI}$ has no deep minimum, 
as shown by Figure~\ref{Fig:hcoup}.
For $M_A = 200$~GeV, the $H^0$ state is much lighter and produces a
much larger contribution to $\sigma_{SI}$.  In this case, the minima of 
$\sigma_{SI}$ are closer to $\pi/2$, near the zeroes of the 
$H^0$-neutralino coupling, as can be seen in Figure~\ref{Fig:Hcoup}.

The values of the electron EDM, to be discussed in the next section, are 
also indicated in Figures~\ref{fig:dd1}--\ref{fig:dd3}. 
Among the direct detection experiments, CDMS excludes the region above 
the line labeled as CDMS 2004.  The lower lines 
indicate the projected sensitivities of future experiments: CDMS 
\cite{Baudis:tv}, ZEPLIN \cite{Cline:pi} and XENON \cite{Aprile}.

\begin{figure*}[htb]
\begin{minipage}[t]{0.49\textwidth}
        \includegraphics[width= \textwidth]{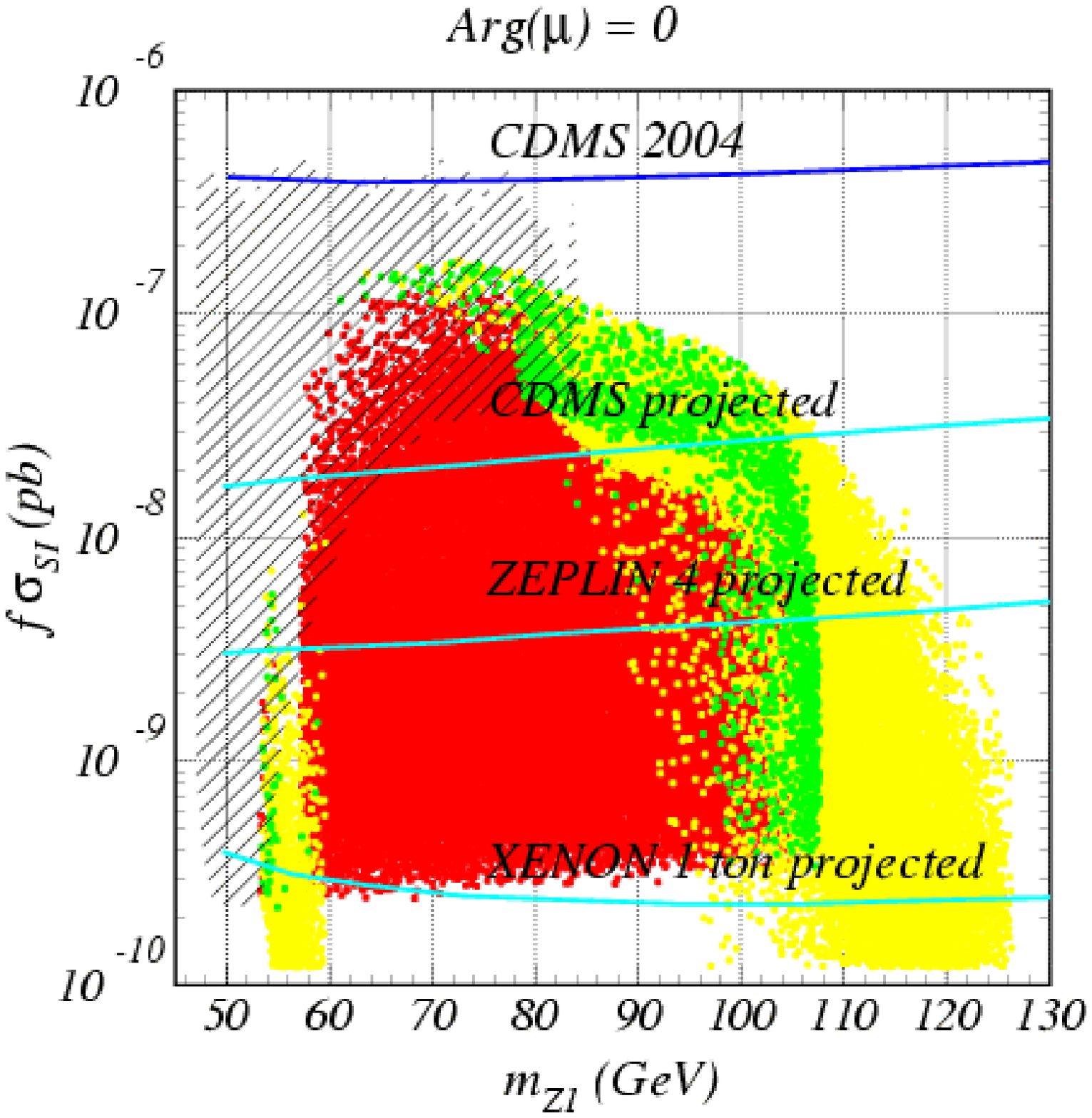}
\end{minipage}
\phantom{aa}
\begin{minipage}[t]{0.49\textwidth}
        \includegraphics[width= \textwidth]{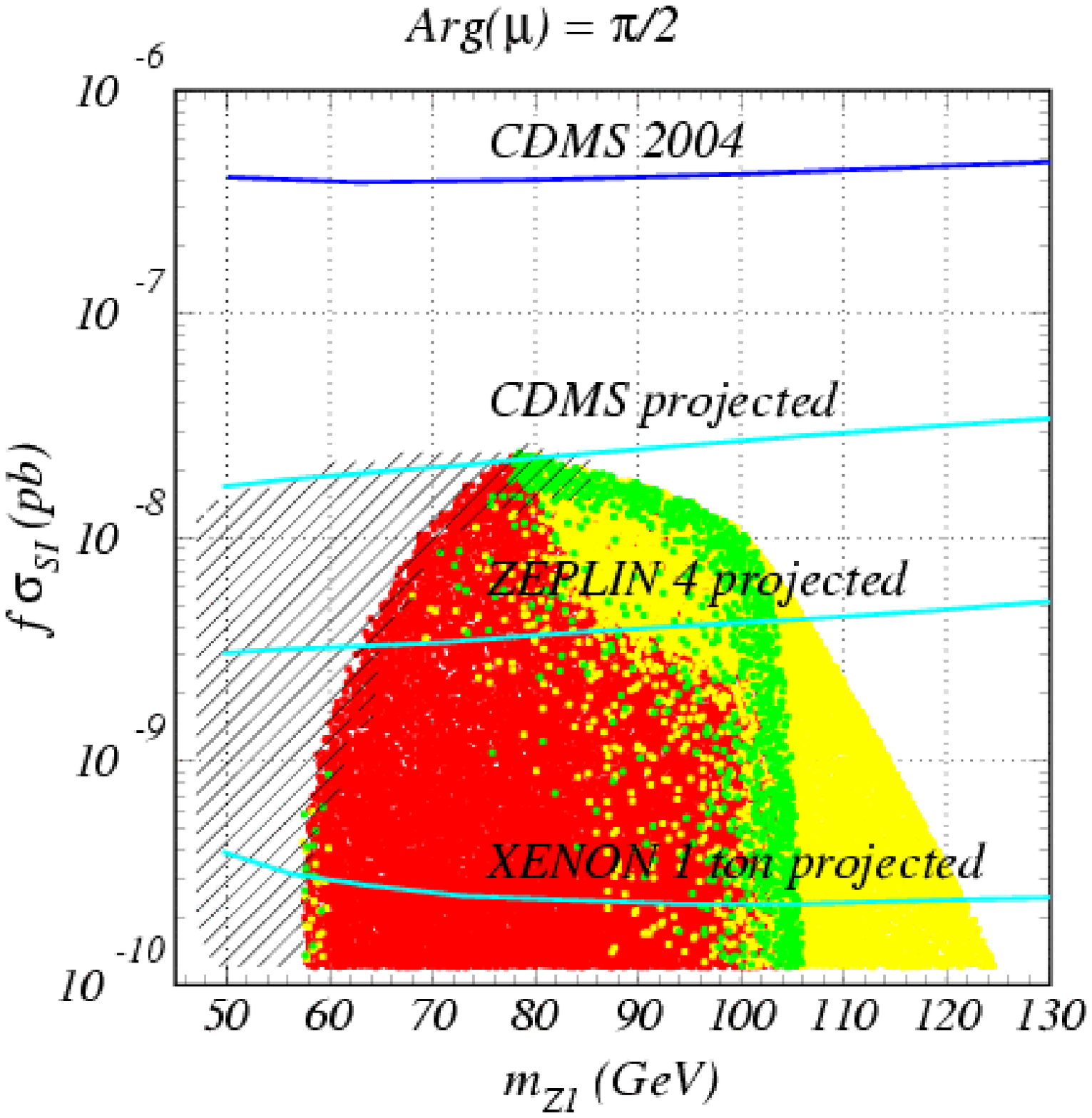}
\end{minipage}
        \caption{Spin independent neutralino-proton elastic 
scattering cross sections as 
a function of the neutralino mass for $Arg(\mu)= 0$ (left) and
$Arg(\mu)= \pi/2$ (right). 
Red (dark gray), green (medium gray) and yellow (light gray) 
dots represent models in which the neutralino density is above, consistent 
or below the 2 $\sigma$ WMAP bounds. 
Hatching indicates the region excluded by chargino searches at LEP. 
The top (blue) solid line represents
the 2004 exclusion limit by CDMS. 
The lower solid (cyan) lines indicate the projected 
sensitivity of CDMS, ZEPLIN and XENON, respectively.}
\label{fig:sigmasivsmz1}  
\end{figure*}

In Figure~\ref{fig:sigmasivsmz1}, we examine the dependence of the direct 
dark matter detection on the phase of $\mu$. In order to do this, we 
conducted a random scan over the following range of MSSM
parameters:
\begin{eqnarray}
&&    -(80 ~{\rm GeV})^2 < m_{\tilde U_3}^2 < 0 , ~~~ 
    100 ~{\rm GeV} < |\mu| < 500 ~{\rm GeV}, ~~~
     50 ~{\rm GeV} < M_1 < 150 ~{\rm GeV}, \nonumber \\
&&  200 ~{\rm GeV} < M_A < 1000 ~{\rm GeV}, ~~~
      5 < \tan\beta < 10.
\label{eq:RandomScanPars}
\end{eqnarray}
The parameters which are not scanned over are fixed as in Section 2.
The result of the scan, projected on the stop mass 
versus neutralino mass plane, is shown by 
Figure \ref{fig:sigmasivsmz1}.  Here we plot $f \sigma_{SI}$ as the function 
of the lightest neutralino mass, where
\begin{eqnarray}
f = \left\{ \begin{array}{ll}
 \Omega_{CDM} h^2/0.095 & {\rm if}~ 0.095 \geq \Omega_{CDM} h^2 \\
 1                      & {\rm if}~ 0.095 < \Omega_{CDM} h^2
\end{array} \right.
\end{eqnarray}
accounts for the diminishing flux of neutralinos with their decreasing 
density \cite{Ellis:2000jd}.\footnote{
The experimental limits for dark matter detection rely on the
standard assumptions of a dark matter flux incident on
the earth, based on the observational evidence that points
to a roughly spherical distribution of dark matter distribution
in the galaxy, and a local dark matter velocity comparable to the
speed of the sun within the galaxy.}  
For models marked by yellow (light gray) 
dots the neutralino relic density is below the 2 $\sigma$ WMAP bound, while 
models represented by green (medium gray) dots comply with WMAP within 2 
$\sigma$.  Models that are above the WMAP value by more than 2 $\sigma$ are 
indicated by red~(dark gray) dots.  The area indicated by hatching is 
excluded by the LEP chargino mass limit of 103.5 GeV.
The top solid (blue) line represents the 2004 exclusion limit by CDMS 
\cite{Akerib:2004fq}.  The lower solid (cyan) lines indicate the 
projected sensitivity of the CDMS~\cite{Baudis:tv}, ZEPLIN~\cite{Cline:pi} 
and XENON~\cite{Aprile:2004ey} experiments. 

The structure of this scatter plot is clear by examining Figures 
\ref{fig:dm1}-\ref{fig:dm3}.  As shown on these plots by the gray direct 
detection contours, the spin-independent cross section, $\sigma_{SI}$,
decreases for increasing values of $|\mu|$. Therefore, the low $\sigma_{SI}$
region in Figure~\ref{fig:sigmasivsmz1} is in one to one correspondence
with the large $|\mu|$ region 
in Figures~\ref{fig:dm1}-\ref{fig:dm3}. 
For  large values of $|\mu|$, 
the lightest neutralino mass is approximately given by $M_1$, hence, 
increasing values of $M_1$ in Figures~\ref{fig:dm1}-\ref{fig:dm3} 
correspond to 
increasing values of  $m_{\widetilde Z_1}$ in 
Figure~\ref{fig:sigmasivsmz1} and the same annihilation 
regions, via  $h^0$ and  $A^0/H^0$ 
resonances, and stop coannihilation regions of
 Figures ~\ref{fig:dm1}-\ref{fig:dm3} 
can be identified in a clear way in Figure~\ref{fig:sigmasivsmz1}.
The LEP excluded, hatched area of $m_{\widetilde \chi_1} < 103.5$ GeV,
preserves its hyperbolic shape for $m_{\widetilde Z_1} < 85$ GeV.

Presently, the region above the (blue) top solid line is excluded by CDMS. 
In the near future, for $Arg(\mu)=0$, CDMS will probe part of the region 
of the parameter space where the WMAP dark matter bound is satisfied.   
In this region, due to their enhanced Higgsino components, neutralinos mainly 
annihilate to gauge bosons or, due to the small mass gap, they coannihilate 
with charginos. The ZEPLIN experiment will start probing the stop-neutralino 
coannihilation region together with the annihilation region enhanced by s-channel 
$A^0$ resonances. Finally, XENON will cover most of the relevant parameter 
space.
Prospects for direct detection of dark matter tend to be worse for large values of
the phase of $\mu$, $Arg(\mu) \simeq \pi/2$.  As seen from 
Figures \ref{fig:dd1}-\ref{fig:dd3}, this phase can lead to cancellations
which suppress the direct detection cross section. 
In the event of such a cancellation, a detector with the sensitivity 
of ZEPLIN is needed to start probing the parameter space, and not 
even XENON will be capable of fully exploring this model.

\section{Constraints on CP Violating Phases\label{edm}}
\subsection{Electron EDM Constraints}

  The MSSM can accommodate many CP violating phases in addition to the 
CKM phase present in the SM.  Such phases, however, are very 
highly constrained by the experimental limits on the electric 
dipole moments~(EDM) of the electron, neutron, and $^{199}$Hg atom.  
Of these, we will focus our attention on the electron EDM 
since it is the best measured, the least plagued by theoretical 
uncertainties, and for the phases relevant to the model under study
gives the strongest constraint. 
The upper bound on the electron EDM comes from measurements
of the EDM of the $^{205}$Tl atom.  
For the phases considered in this work and in the absence of Higgs mixing,
the CP-odd electron-neutron operator studied in~\cite{Pilaftsis:2002fe} 
vanishes, and the $^{205}$Tl EDM is due almost entirely to the electron EDM.
This translates into a limit on the electron EDM of~\cite{Regan:2002ta}
\be
|d_e| < 1.6\times 10^{-27}\;e\,cm,
\label{eedm}
\ee
at $90\%$~CL.  

  In the MSSM, the leading order contributions to the electron EDM 
come from one-loop diagrams containing an intermediate
selectron or sneutrino.  For $\mathcal{O}(1)$ phases, 
these loops generate an EDM well above the experimental
limit unless these sfermions are taken to be quite heavy,
$m_{\tilde{f}}\gtrsim 10$~TeV~\cite{Abel:2001vy}.
The neutron and $^{199}$Hg EDM constraints require that 
the other first and second generation sfermions be very heavy
as well.  This feature arises in several models considered in the 
literature ~\cite{Cohen:1996vb,Feng:1999zg,Arkani-Hamed:2004fb,
Carena:2004ha}. 
Such large first and second generation sfermion masses 
present no problem for EWBG since they couple very weakly 
to the Higgs bosons, and have only a minor effect on the final 
CP asymmetry~\cite{improved}.  With respect to EWBG, 
a much more dangerous contribution arises at two-loops.

  At the two-loop order there are relevant contributions to the electron
EDM from loops containing intermediate charginos and Higgs bosons.
Since EWBG demands that the charginos be fairly light,
$m_{\chi} \lesssim 500$~GeV, these contributions cannot be
suppressed by taking large chargino masses.  On the other hand,
these terms can be reduced by taking large $M_A$ or small $\tan\beta$.
The phase associated with this contribution
comes primarily from the chargino mass matrix, which is the same
phase that generates the baryon asymmetry, and lower values of $M_A$ can 
enhance the baryon asymmetry.  Consequently, the 
electron EDM bound presents a particularly severe constraint
on EWBG within the MSSM.  

  We have examined whether it is possible for EWBG to generate
the observed baryon asymmetry while obeying the 
electron EDM bounds.  The two-loop contributions to the electron 
EDM due to intermediate charginos and Higgs were calculated following
~\cite{Chang:2002ex,Pilaftsis:2002fe}.  The method of~\cite{Carena:2002ss}
was used to calculate the baryon asymmetry generated by EWBG. 
In our analysis, we have fixed 
$M_2 =200$~GeV, and varied $\mu$, $Arg(\mu)$, 
$\tan\beta$, and $M_A$.  
We also assume a bubble wall velocity of $v_w = 0.05$ and 
a wall width of $L_w = 20/T$.  Both of these values are fairly
typical, and tend to maximize the baryon asymmetry generated 
in the phase transition.  
  
\begin{figure*}[htb]
\centerline{
        \includegraphics[width=0.55\textwidth]{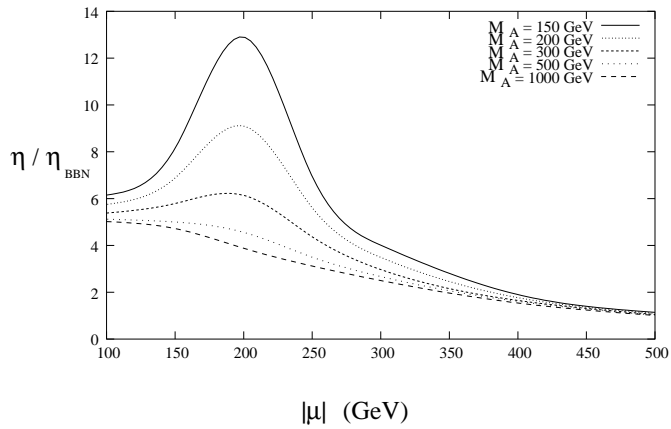}}
        \caption{Baryon asymmetry generated by  EWBG relative to that 
required by big-bang nucleosynthesis for $M_2 = 200$, $\tan\beta = 5$,
and $\sin(Arg(\mu)) = 1$.}
\label{fig:ewbg}        
\end{figure*}

  The dependence of the baryon asymmetry (relative to the value
needed for big-bang nucleosynthesis~(BBN)), $\eta/\eta_{BBN}$, 
on $|\mu|$ and $M_A$ is illustrated in Figure~\ref{fig:ewbg}.
In this plot, we have taken the phase to be maximal, 
$\sin(Arg(\mu)) = 1$, and have set $\tan\beta = 5$.  
For other values of these parameters, the baryon asymmetry
scales with $\sin(Arg(\mu))$ and (approximately) with $\sin 2\beta$.
There are two main contributions from the CP violating currents of
charginos and neutralinos to the baryon asymmetry in the MSSM.  
The first is proportional to the change in 
$\beta$ going from the symmetric phase to the broken phase 
and exhibits a resonance at $M_2 = |\mu|$,
but is highly suppressed for large values of $M_A$.
The second contribution is independent of $M_A$, and falls 
off smoothly as $|\mu|$ becomes large.  Both contributions
go to zero as $M_2$ becomes large.

\begin{figure*}[hbt!]
\begin{minipage}[t]{0.47\textwidth}
        \includegraphics[width = \textwidth]{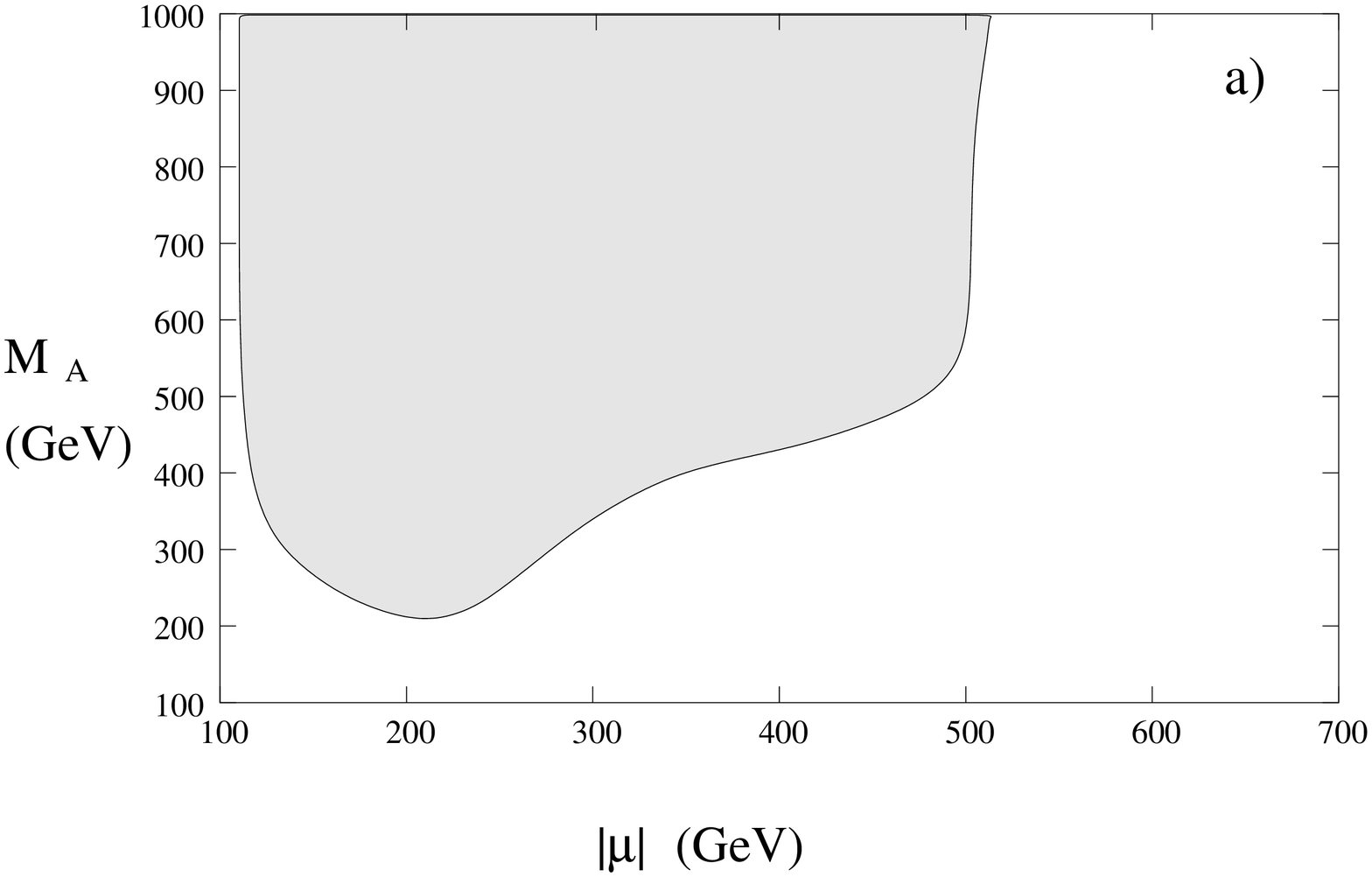}
\end{minipage}
\phantom{aa}
\begin{minipage}[t]{0.47\textwidth}
        \includegraphics[width = \textwidth]{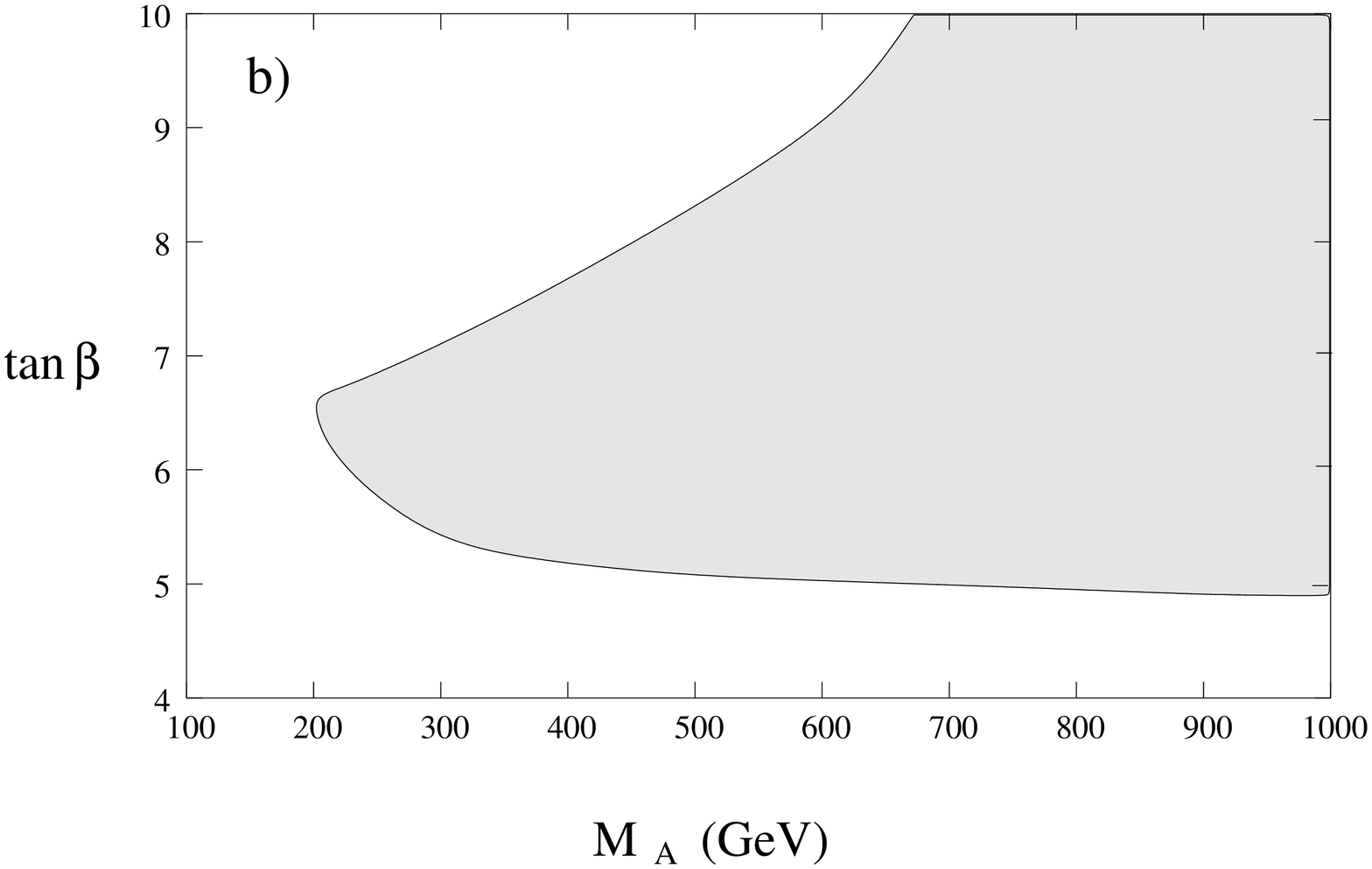}
\end{minipage}
\caption{Parameter regions consistent with EWBG and the 
electron EDM limit.  In these plots, we have taken 
$M_2=200~\mbox{GeV}$ and varied $Arg(\mu)$ over 
the interval $[0,\pi]$.}
\label{muma}
\end{figure*}

  Figures~\ref{muma} a) and \ref{muma} b) show the regions in the 
$|\mu|\!-\!M_A$ and $M_A\!-\!\tan\beta$ planes consistent with both
EWBG and the experimental bound on the electron EDM. 
Here, we have scanned over the ranges
\be
3 < \tan\beta < 10,\:\:\: 100~\mbox{GeV} < M_A < 1000~\mbox{GeV},\:\:\: 
100~\mbox{GeV} < |\mu| < 1000~\mbox{GeV},
\ee
with $M_2 = 200$~GeV and the rest of the parameters as in Section~2.
In Figure~\ref{muma}a) we see that in the allowed region,
$|\mu|$ is confined to the range $110\lesssim|\mu|\lesssim 
550$~GeV, while $M_A$ must be greater than about 200~GeV. 
The limits on $|\mu|$ are due to the effect of this parameter 
on the chargino mass.  For $|\mu| \lesssim 110$~GeV, the lighter
chargino has mass below the experimental bound, 
$m_{\chi_1}\gtrsim 103.5$~GeV~\cite{LEPSUSYW1}, 
while for large $|\mu|$, EWBG becomes
less efficient.  The lower bound on $M_A$ arises for two reasons.
For small $M_A$ the two-loop contribution to the electron EDM
is enhanced.  At the same time the mass of the lightest Higgs is suppressed.
The effect of the Higgs mass constraint can also be seen
in Figure~\ref{muma}b), in which this bound results in a lower 
limit on $\tan\beta$.  The allowed region is 
cut off for larger values of $\tan\beta$ since this tends to 
enhance the two-loop contributions to the electron EDM.

\begin{figure*}[hbt!]
\begin{minipage}[t]{0.47\textwidth}
        \includegraphics[width = \textwidth]{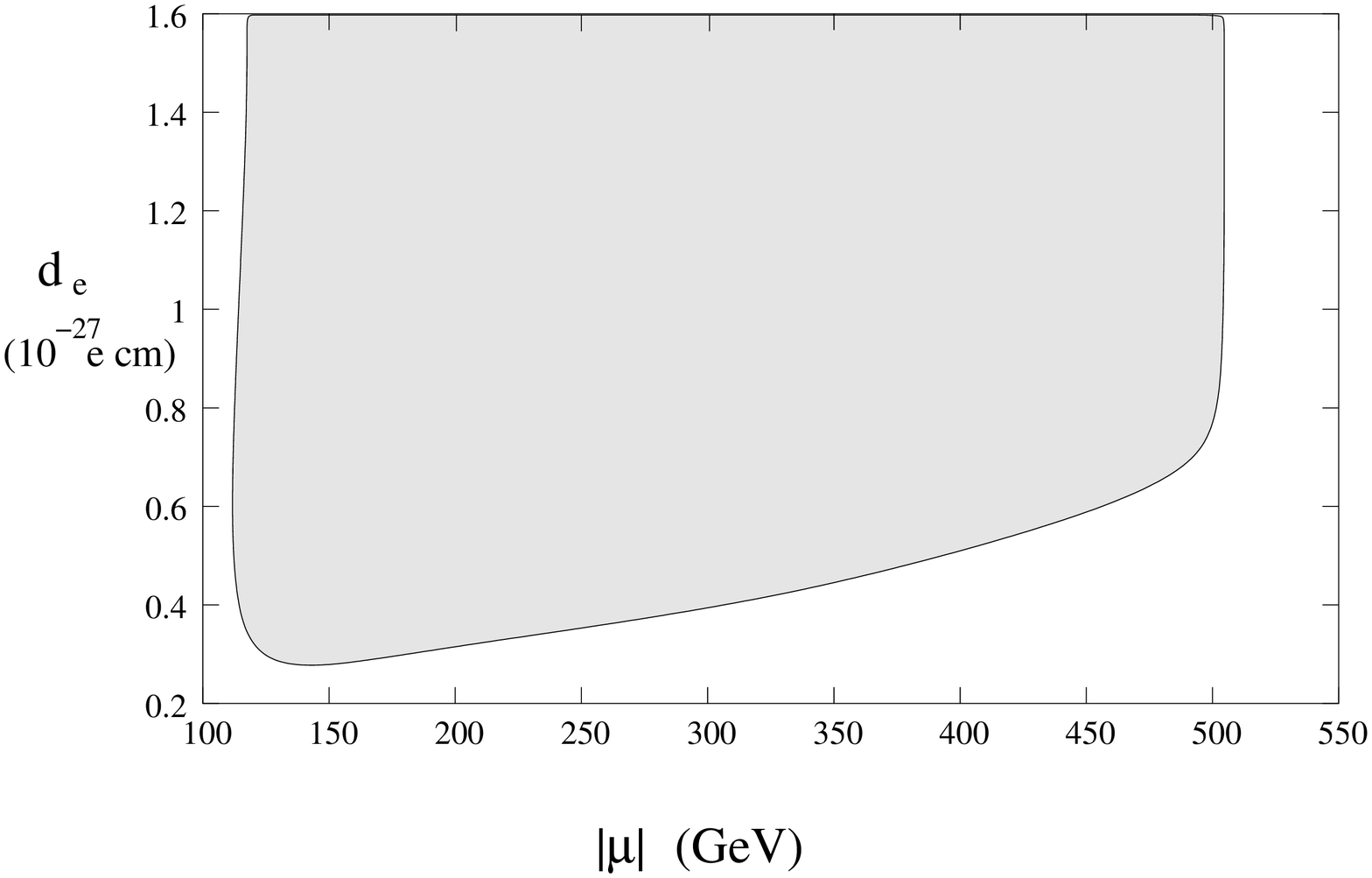}
\end{minipage}
\phantom{aa}
\begin{minipage}[t]{0.47\textwidth}
        \includegraphics[width = \textwidth]{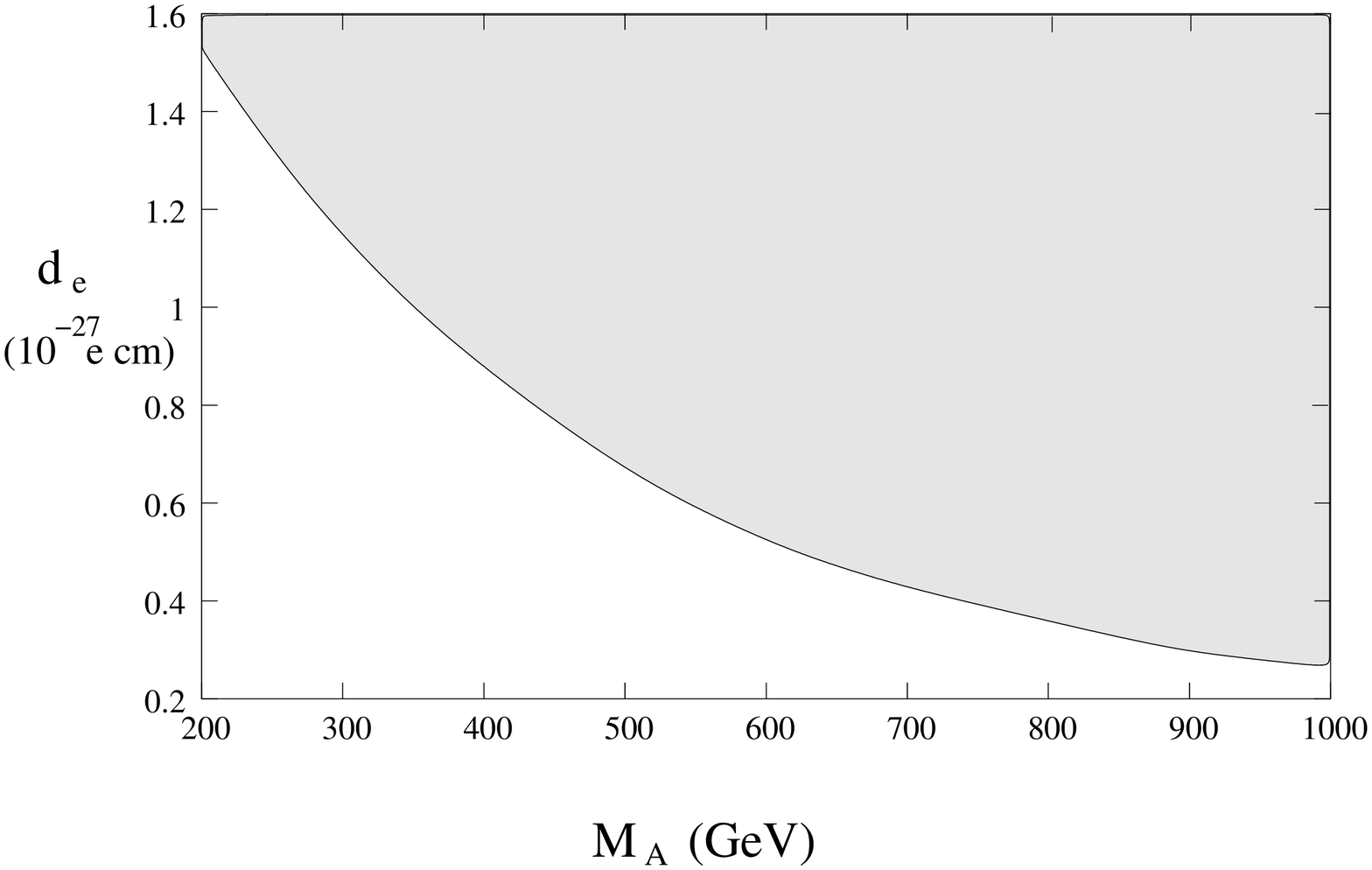}
\end{minipage}
\caption{Range of values of the electron EDM for parameter sets
consistent with EWBG.}
\label{made}
\end{figure*}

  From Figure~\ref{muma}, we see that it is possible to generate
the baryon asymmetry via EWBG in the MSSM while satisfying
the experimental constraints on the electron EDM and the 
mass of the lightest Higgs boson.  Although this is reassuring,
the EWBG scenario is still very strongly constrained by the electron EDM.  
This can be seen in Figure~\ref{made}, which shows the 
range of values of $d_e$ obtained in our scan that are consistent
with EWBG, the current electron EDM bound, and the Higgs mass limit.
For $M_A < 1000$~GeV, an order of magnitude improvement 
of the electron EDM bound, $|d_e| < 0.2\times 10^{-27}\;e\,cm$, 
will be sufficient to test this baryogenesis mechanism within the MSSM.  
However, we should also point out that the calculation of the baryon 
asymmetry from EWBG has $\mathcal{O}(1)$ uncertainties associated with 
the values of the bubble parameters, the wall velocity, and the derivative expansion 
used to derive the diffusion equations. Hence, the limits on EWBG presented 
here may be somewhat more (or less) severe than they really are.
Furthermore, we have not considered the possibility of fortuitous 
cancellations between different EDM contributions, for instance between the 
one-loop and two-loop terms (for lighter sfermions), 
which could further reduce the value of the electron EDM.

\subsection{Constraints from ${\rm BR}(b \to s \gamma)$}
\label{bsg}

The presence of a light stop, light charginos, and a light charged Higgs 
boson may induce relevant effects on flavour changing neutral currents
associated with the bottom quark \cite{Murayama:2002xk}.
 One of the most sensitive experimental
measurements of such effects is the branching ratio of the decay of
a bottom quark into a strange quark and a 
photon~\cite{Bertolini:1990if}--\cite{Carena:2000uj}. 
A realistic calculation
of these effects, however, cannot be performed without knowledge of the
flavour sector of the theory. Even for the large values of the
bottom squark masses we consider in this work, of order of a few TeV, the
contributions coming from the interchange of gluinos and down squarks
may be as large as the ones coming from the stop--chargino 
loops~\cite{Masiero}.

  In the following, we shall present the results for the branching
ratio of this rare decay, assuming that the only relevant contributions
beyond the SM ones are those associated with the charged Higgs and
stop--chargino loops. While the former tend to increase the
$BR(b \to s \gamma)$ compared to the SM value, the latter has a non-trivial 
dependence on the CP violating phase. 
The experimental value of $BR(b \to s \gamma)$ is given by~\cite{bsgaexp},
\be
{\rm BR}(b \to s \gamma) = (3.54^{+0.30}_{-0.28}) \times 10^{-4}
\ee

\begin{figure*}[hbt!]
\begin{minipage}[t]{0.47\textwidth}
        \includegraphics[width = \textwidth]{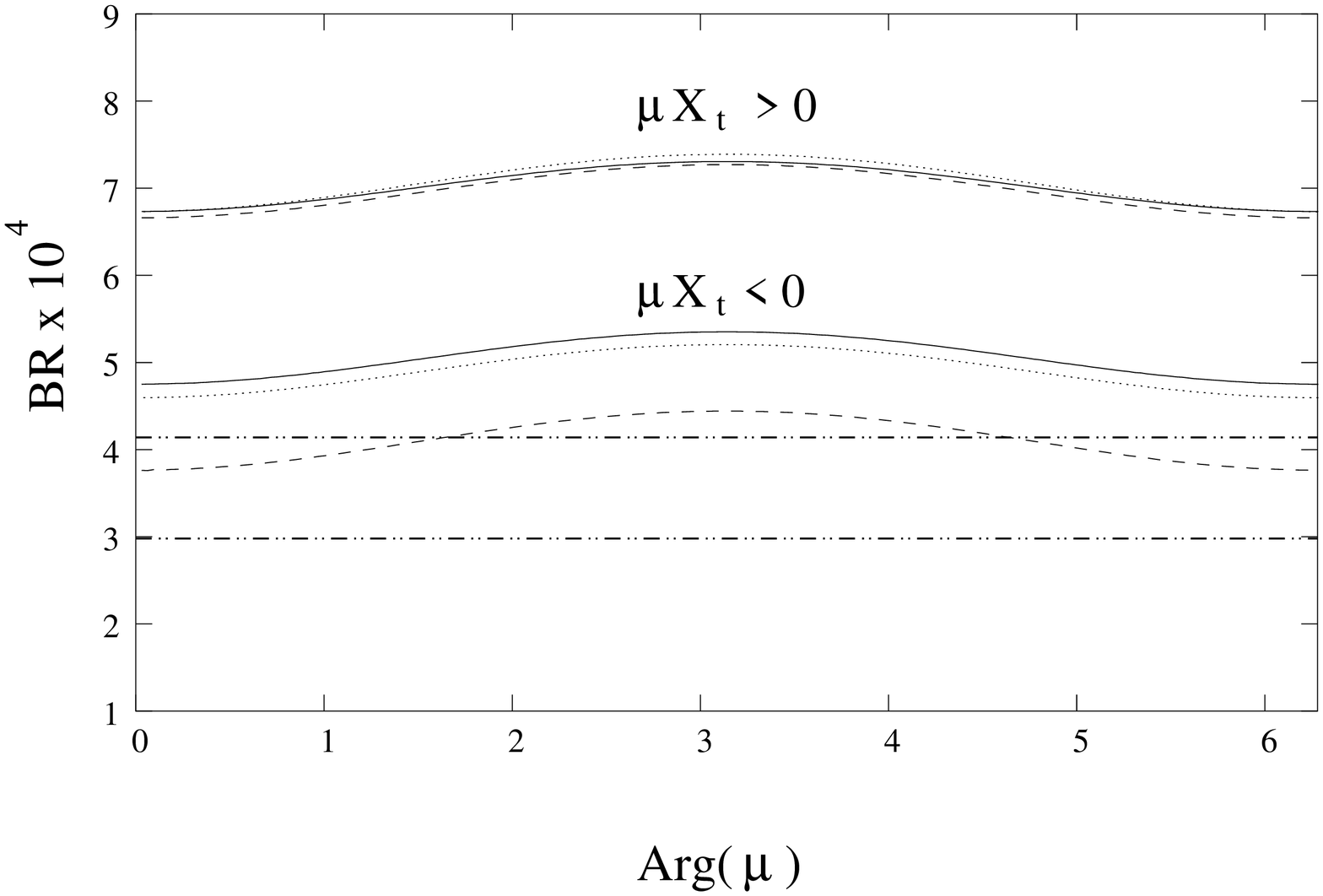}
\end{minipage}
\phantom{aa}
\begin{minipage}[t]{0.47\textwidth}
        \includegraphics[width = \textwidth]{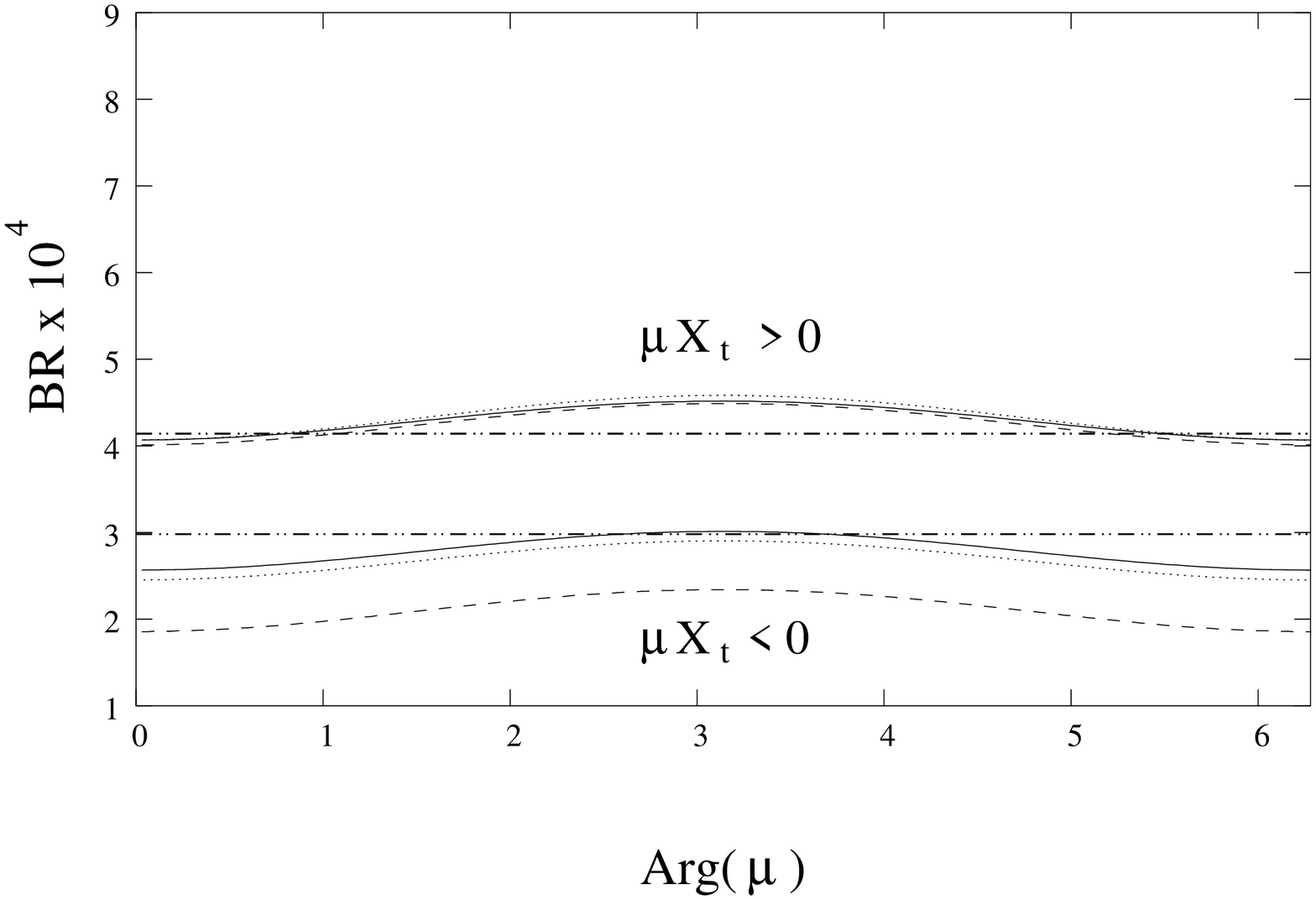}
\end{minipage}
\caption{${\rm BR}(b\to s\gamma)$ as a function of $Arg(\mu)$ for
values of the CP-odd Higgs mass $M_A = 200$~GeV (left), and  
$M_A = 1000$~GeV (right).  The stop parameters were chosen as in section 2, 
and the chargino and neutralino mass parameters are given by $(|\mu|,M_1) =$ 
(350, 110)~GeV (solid lines), (175, 110)~GeV (dashed lines), 
and (300, 60)~GeV (dotted lines).  The dot-dashed bands represent 
the present experimental range at the 2 $\sigma$ level.} 
\label{fig:bsga}
\end{figure*}

In figure~\ref{fig:bsga} we display the value of ${\rm BR}(b \to s \gamma)$
as a function of the phase of the Higgsino mass parameter, $\mu$,
for $M_A = 200,\:1000$~GeV. 
The stop sector parameters have been chosen as in section~2, while the chargino and
neutralino mass parameters are taken to be $(|\mu|,M_1) =$
(300, 60)~GeV (solid lines), (350,110)~GeV (dashed lines),  
and (175,110)~GeV (dotted lines). 

  As is apparent from the figure, in the absence of other sources of 
flavour violation, a light CP-odd Higgs scalar with mass of about 
200~GeV is highly restricted by $BR(b \to s \gamma)$.  
Negative values of $\mu\,X_t$, where 
$X_t = A_t - \mu^*/\tan\beta$, are necessary to 
keep the predicted branching ratio close to the experimentally allowed 
range.\footnote{Recall that if the phases originate from a common
gaugino phase and a U(1)$_R$ transformation 
is used to transfer this phase to $\mu$
and $A_t$, the product $\mu\,A_t$ remains real but can have either sign. 
See Eq.(\ref{Eq:u1r}).}
This is due to a cancellation between the charged Higgs and the
squark--chargino contributions to the branching ratio when $\mu\,X_t$ is negative.
Otherwise these contributions interfere constructively 
with each other and with the SM contribution.
For both signs of $\mu\,X_t$, the branching ratio is largest when
$Arg(\mu) = \pi$ and smallest for $Arg(\mu) = 0$.
Since the branching ratio tends to be somewhat high for $M_A \sim 200$~GeV,
even with $\mu\,X_t<0$, small values of $Arg(\mu)$ are preferred in this case.

  Larger values of the CP-odd Higgs mass are consistent with the measured value
of ${\rm BR}(b \to s \gamma)$ over a wide range of values of 
$M_1$, $\mu$, and $Arg(\mu)$.
For moderately large values, $M_A\lesssim 1000$~GeV, negative $\mu\,X_t < 0$ 
is preferred.  For $M_A \gtrsim 1000$~GeV, the charged Higgs contribution 
decouples leaving only the stop-chargino corrections.  
These corrections tend to give a branching ratio 
that is near the upper part ($\mu\,X_t > 0$) or lower part ($\mu\,X_t < 0$) of
the experimentally allowed range for $|X_t| = 700$~GeV, as we have considered here.
Thus, smaller $Arg(\mu)$ is preferred for $\mu\,X_t > 0$, while $Arg(\mu)\sim\pi$
is preferred for $\mu\,X_t < 0$.
The chargino corrections can be reduced in size by taking 
slightly smaller values of $|X_t|$, or by invoking small flavour violation 
effects in the down squark sector.

\section{Conclusions\label{concl}}

Electroweak baryogenesis provides a mechanism for the generation
of the baryon asymmetry that relies only on physics at the
weak scale.  It is therefore testable at high energy physics
facilities in the near future. In a previous work, we showed
that satisfactory dark matter abundance may be obtained
in the presence of a light stop like the one consistent with
electroweak baryogenesis, and analyzed the impact of the
allowed parameter space for stop searches at hadron colliders.
No CP violating effects were considered.

In this work, we have analyzed the effect of CP violating
phases, as required for EWBG, in conjunction with a light stop,
with mass below the top quark mass, and
a light Higgs with mass below 120~GeV.
 We have shown that these phases have only a minor
impact on the stop--neutralino parameter space leading
to a consistent relic density. Large phases, however, 
have a relevant impact on direct dark matter detection 
rates and  induce large corrections to the electron
electric dipole moment. 

  We have also shown that, for the phases necessary to obtain an
acceptable baryon asymmetry and in the limit of
heavy squarks, of order a few TeV, the predicted values of 
the electron electric dipole moment tend to lie within
an order of magnitude below the reach of  the present bounds. Even in the
case of very heavy squarks, two-loop induced EDM's become
relevant. Assuming no cancellations between one- and two-loop
corrections, one can obtain strong bounds on the allowed 
parameter space: While small
values of $\tan\beta$ are excluded since they
lead to unacceptably small values of the Higgs mass, large
values of $\tan\beta$ tend to lead to unacceptably large
values of the electron EDM or small values of the 
baryon asymmetry. 
On the other hand, for moderate values
of $\tan\beta \simeq 7$, the Higgs boson mass may be large
enough to evade the LEP bounds, even for values of $M_A$
as small as 200 GeV. For this particular value, and for
$|\mu| \simeq M_2$, the baryon asymmetry may be large
enough to be consistent with observations,  even for small values
 of the phases, of order 0.1, for which the EDM's
are consistent with the present experimental bounds.

In the above, we have not discussed the prospects of 
stop searches at hadron and lepton colliders.
As discussed in Ref.~\cite{Balazs:2004bu},  
stop searches become very challenging in the
region where stop-neutralino coannihilation becomes
relevant, both at the LHC and the Tevatron collider~\cite{joekon},
due to the small mass difference 
between the stop and the neutralino. 
An acceptable dark matter density may
be obtained for mass differences as small as 20~GeV, for which
the charm particles proceeding from the stop decay are soft,
making the stop detection difficult. 
As shown in Figures~1--3, the presence of CP violating
phases doesn't affect this result.

The linear collider signatures of MSSM Baryogenesis have been discussed in 
Ref.\cite{Murayama:2002xk}.  A  linear collider 
represents the best possibility for confirming this scenario since it provides
the opportunity of performing precise measurements of the chargino system 
and hence the possibility of observing  a non-zero phase of the $\mu$ 
parameter~\cite{Choi:1998fh}. Precise measurements of the stop system also
become easier at a linear electron-positron collider~\cite{stopsLC}.
For instance, the LEP collider was able to set limits on the stops
even for a mass difference with the neutralino of about 1~GeV. Preliminary
studies of stop searches at the linear collider~\cite{Ayres} show
that a 500~GeV ILC may be able to detect a light stop for mass
differences as small as a few GeV. As said before,
in the region of
parameters where stop-neutralino coannihilation 
leads to a value of the relic density consistent with
experimental results, the stop-neutralino mass difference is 
never much smaller than
20~GeV, and hence an ILC will be able to explore this region 
efficiently

In summary, the requirement of a consistent generation of
baryonic and dark matter in the MSSM leads to
a well-defined scenario, where,
apart from a light stop and a light Higgs boson, one
has light neutralinos and charginos, sizeable
CP violating phases, and moderate values of 
$5 \simlt \tan\beta \simlt 10$.
All these properties will be tested by the Tevatron, the LHC and a 
prospective ILC,
as well as through direct dark-matter detection 
experiments in the near future. The first  tests of this 
scenario will come from electron EDM measurements, stop searches
at the Tevatron and Higgs searches at the LHC 
within the next few years.

\section*{Acknowledgements}

M.C. and C.W. would like to thank M. Quiros and M. Seco
for very useful discussions.
We also gratefully acknowledge the use of {\it Jazz}, a 350-node
computing cluster operated by the Mathematics and Computer Science Division
at ANL as part of its Laboratory Computing Resource Center.
Work at ANL is supported in part by the US DOE, Div.\ of HEP, Contract 
W-31-109-ENG-38. Fermilab is operated by Universities Research Association 
Inc. under contract no. DE-AC02-76CH02000 with the DOE. 


\end{document}